\documentclass[fleqn,usenatbib]{mnras}

\usepackage{newtxtext,newtxmath}

\usepackage[T1]{fontenc}

\DeclareRobustCommand{\VAN}[3]{#2}
\let\VANthebibliography\thebibliography
\def\thebibliography{\DeclareRobustCommand{\VAN}[3]{##3}\VANthebibliography}

\usepackage{graphicx}	
\usepackage[normalem]{ulem}

\usepackage{tabularx}
\usepackage{multirow}
\usepackage{xcolor}

\newcommand{\beq}{\begin{equation}}
\newcommand{\eeq}{\end{equation}}

\def\Mpc{\, h^{-1} \, {\rm Mpc}}

\def\Ms{\, h^{-1} \, M_\odot}

\title[Galaxy-halo connection with ML]{Modeling the galaxy-halo connection with machine learning}

 \author[Delgado et al.]{Ana Maria Delgado,$^{1}$\thanks{E-mail: ana\_maria.delgado@cfa.harvard.edu (AMD)}
Digvijay Wadekar,$^{2,3}$
Boryana Hadzhiyska,$^{1}$
Sownak Bose,$^{1,7}$
Lars Hernquist,$^{1}$\newauthor
Shirley Ho$^{2,4,5,6}$\\
\\
$^{1}$Center for Astrophysics | Harvard \& Smithsonian, 60 Garden Street, Cambridge, MA 02138, USA\\
$^{2}$Center for Cosmology and Particle Physics, Department of Physics,
New York University, New York, NY 10003, USA\\
$^{3}$School of Natural Sciences, Institute for Advanced Study, Princeton, NJ 08540, USA\\
$^{4}$Center for Computational Astrophysics, Flatiron Institute, 162 5th Ave, New York, NY 10010, USA\\
$^{5}$Department of Astrophysical Sciences, Princeton University, Peyton Hall, Princeton NJ 08544-0010, USA\\
$^{6}$Department of Physics, Carnegie Mellon University, Pittsburgh, PA 15217, USA\\
$^{7}$Institute for Computational Cosmology, Department of Physics, Durham University, Durham DH1 3LE, UK
}

\date{Accepted XXX. Received YYY; in original form ZZZ}

\pubyear{2021}

\begin{document}
\label{firstpage}
\pagerange{\pageref{firstpage}--\pageref{lastpage}}
\maketitle

\begin{abstract}
To extract information from the clustering of galaxies on non-linear scales, we need to model the connection between galaxies and halos accurately and in a flexible manner. Standard halo occupation distribution (HOD) models make the assumption that the galaxy occupation in a halo is a function of only its mass, however, in reality, the occupation can depend on various other parameters including halo concentration, assembly history, environment, spin, etc.
Using the IllustrisTNG hydrodynamical simulation as our target, we show that machine learning tools can be used to capture this high-dimensional dependence and provide more accurate galaxy occupation models. Specifically, we use a random forest regressor to identify which secondary halo parameters best model the galaxy-halo connection and symbolic regression to augment the standard HOD model with simple equations capturing the dependence on those parameters, namely the local environmental overdensity and shear, at the location of a halo. This not only provides insights into the galaxy-formation relationship but, more importantly, improves the clustering statistics of the modeled galaxies significantly.
Our approach demonstrates that machine learning tools can help us better understand and model the galaxy-halo connection, and are therefore useful for galaxy formation and cosmology studies from upcoming galaxy surveys.

\end{abstract}Submit to MNRAS


\section{Introduction}

As we aspire to solve some of astronomy's toughest challenges, such as constraining  cosmological parameters, measuring neutrino masses or determining the nature of dark energy, the community has planned galaxy surveys of unparalleled depth (e.g., Euclid \citep{2020A&A...643A..70T}, DESI \citep{desicollaboration2016desi}. These surveys will provide us with far more accurate and precise measurements of the large scale distribution of matter and will primarily use galaxy clustering statistics to study galaxy formation and cosmology. We therefore want to make sure that our statistical models and systematic uncertainties are robust enough to handle the precision needed to achieve an accuracy of order 1\%.

An important step in fully realising the statistical power of future surveys is to achieve a precise understanding of the galaxy-halo connection. One such example is the standard halo occupation distribution model (HOD) \citep{PeaSmi00, Sel00, Sco00, BerWei02}, which predicts the number of galaxies that reside within a dark matter halo, and in its simplest form depends only on halo mass. We can use results from the HOD model as weights assigned to halos for statistics in order to predict the clustering of galaxies. However, the clustering of halos does not trace the distribution of matter exactly but is instead ‘biased’ relative to it in a manner that depends on properties of the halo beyond simply its mass (so-called ‘halo assembly bias’). Furthermore, the galaxy occupancy per halo has also been found to depend on more than just halo mass \citep{2001MNRAS.328...64N, 2002ApJ...571..172Z}. Thus, in order to recover the true clustering bias of galaxies (as predicted, say, by hydrodynamical simulations) relative to the underlying matter distribution, we need to also assign galaxy occupation based on secondary halo parameters other than halo mass.

Several studies have tried to incorporate secondary halo parameters into an HOD framework to account for the effects of assembly bias. Halo concentration, for example, has been a popular secondary parameter \cite{CroGaoWhi07,VakHah19, KobNisTak20, WecTin18,ParKov15}, although numerous recent studies have shown that the environment of the halos plays a significant role in determining the galaxy distribution \citep{HadBosEis20b,HadBosEis20, Hadzhiyska_2021, YuaHadBos20, McEWei18, XuZeh20, WibSalWei19, SalWibWei20, AbbShe07,PujGaz14}. There also have been multiple prescriptions for including the halo environment in the HOD model \citep{McEWei18, XuZeh20, WibSalWei19, SalWibWei20,YuaHadBos20, HadTac20,SalZu20}.

High-resolution hydrodynamical simulations such as IllustrisTNG (TNG) \citep{Springel_2010, 2018MNRAS.475..624N} give us a plausible prediction of what the galaxy-halo connection should be and can therefore be used as a testing ground for theory. By identifying observables that will best allow us to model galaxy assembly bias we may improve upon our current models and test their statistical power against hydrodynamical simulations.

We have two primary goals in this work:\\
\textit{(i)} Determine which secondary halo properties, in addition to halo mass, best model the galaxy-halo connection. Because there is no first-principles explanation as to which properties to use, we model the number of galaxies using machine learning by approximating the function
\beq
N_\mathrm{galaxies} = f(M_\mathrm{halo}, \{i_\mathrm{halo}\}) \, ,
\label{eq:intro}\eeq
where $M_\mathrm{halo}$ is the total mass of the halo, and $\{i_\mathrm{halo}\}$ is the set of various secondary halo properties: the overdensity and anisotropy of its environment at various scales, its concentration, spin, velocity dispersion, and so on. The high dimensionality of the input space makes this a complex and challenging problem. Correlations between different input parameters (i.e., the concentration of a halo is related to its assembly history and its environment) further add to the difficulty.\\
\textit{(ii)} Augment the standard HOD model with simple equations that incorporate the effects of the secondary halo properties. As central and satellite galaxies can have a different dependence on halo properties, we train over models separately for centrals and satellites throughout this paper. It is worth noting that our approach is similar to \cite{WadVil20b} who used machine learning to model the neutral hydrogen content of the halo as a function of halo mass and secondary properties.

The layout of the paper is as follows: In section \ref{methods}, we describe the simulations, our benchmark model, how we used machine learning algorithms to obtain augmented models, the summary statistics used and a description of the secondary halo properties considered in this work. In section \ref{results} we present our results. Section \ref{discussion} provides discussion on how our work compares to previous studies and describes the limitations of our methods. In section \ref{conclusions} we summarize our findings and conclude.

\section{Methods}
\label{methods}

\subsection{IllustrisTNG simulation}
\label{Method:TNG}

\textit{The Next Generation Illustris}, IllustrisTNG (hereafter, TNG) \citep{2018MNRAS.473.4077P,  2018MNRAS.475..676S, 2018MNRAS.475..624N, 2018MNRAS.477.1206N, 2018MNRAS.480.5113M, 2019ComAC...6....2N, 2019MNRAS.490.3234N, 2019MNRAS.490.3196P}, are cosmological, hydrodynamical simulations run with the AREPO code \citep{Springel_2010,Weinberger2020}, which utilizes a hybrid tree/particle-mesh scheme to solve for gravitational interactions of dark matter particles and an unstructured, moving mesh to solve the equations of hydrodynamics. Compared to the galaxy formation model of its predecessor, \textit{Illustris} \citep{2014MNRAS.444.1518V, 2014Natur.509..177V, 2014MNRAS.445..175G}, the model in TNG has updated implementations of AGN feedback \citep{2017MNRAS.465.3291W} and galactic winds \citep{2018MNRAS.473.4077P}, and incorporates magnetic fields \citep{Pakmor_2014}. The TNG suite consists of three simulation volumes: TNG50, TNG100 and TNG300 each run at three different resolutions. In this work we use the TNG300-1 simulation, a periodic box of length $L_{\mathrm{box}}=205 h^{-1}\mathrm{Mpc} \approx 300 \mathrm{Mpc}$, containing $2\times2500^3$ resolution elements with a mass resolution of $7.6\times10^6 h^{-1}M_{\odot}$ for baryons and $4.0\times10^7 h^{-1}M_{\odot}$ for dark matter.

The initial conditions of the TNG suite were generated at $z=127$ and assume $\textit{Planck}$ parameters \citep{2016A&A...594A..13P}. Each hydrodynamical simulation (henceforth referred to as FP for `full-physics') also has a counterpart generated from the same initial conditions but evolved with dark matter only (\textit{N}-body, DMO). 

Halos in TNG are identified using the `friends-of-friends' (FOF) algorithm which forms groups by connecting together dark matter particles separated by at most 20\% of the mean interparticle separation. Subhalos are identified using the SUBFIND algorithm, which requires that each subhalo contain at least 20 dark matter particles that are gravitationally bound. A galaxy is defined as the constituent baryonic particles (those which make up stars, gas and black holes) associated with the subhalo. 

For this work, we take advantage of the initial-conditions matched between FP and DMO simulations to create bijective matches of halos between the two runs, as outlined in \cite{HadBosEis20}. This allows us to mimic the standard implementation of the Halo Occupation Distribution (HOD) for the DMO run by populating its halos with galaxy occupation numbers per halo mass bin derived from the FP run. Specifically, we use the TNG300-1 DMO simulation populated with its bijectively matched halos (henceforth TNG300-matched or TNG300). 

\subsection{Benchmark model: Halo occupation distribution (HOD)}

\begin{figure}
\includegraphics[width=\columnwidth]{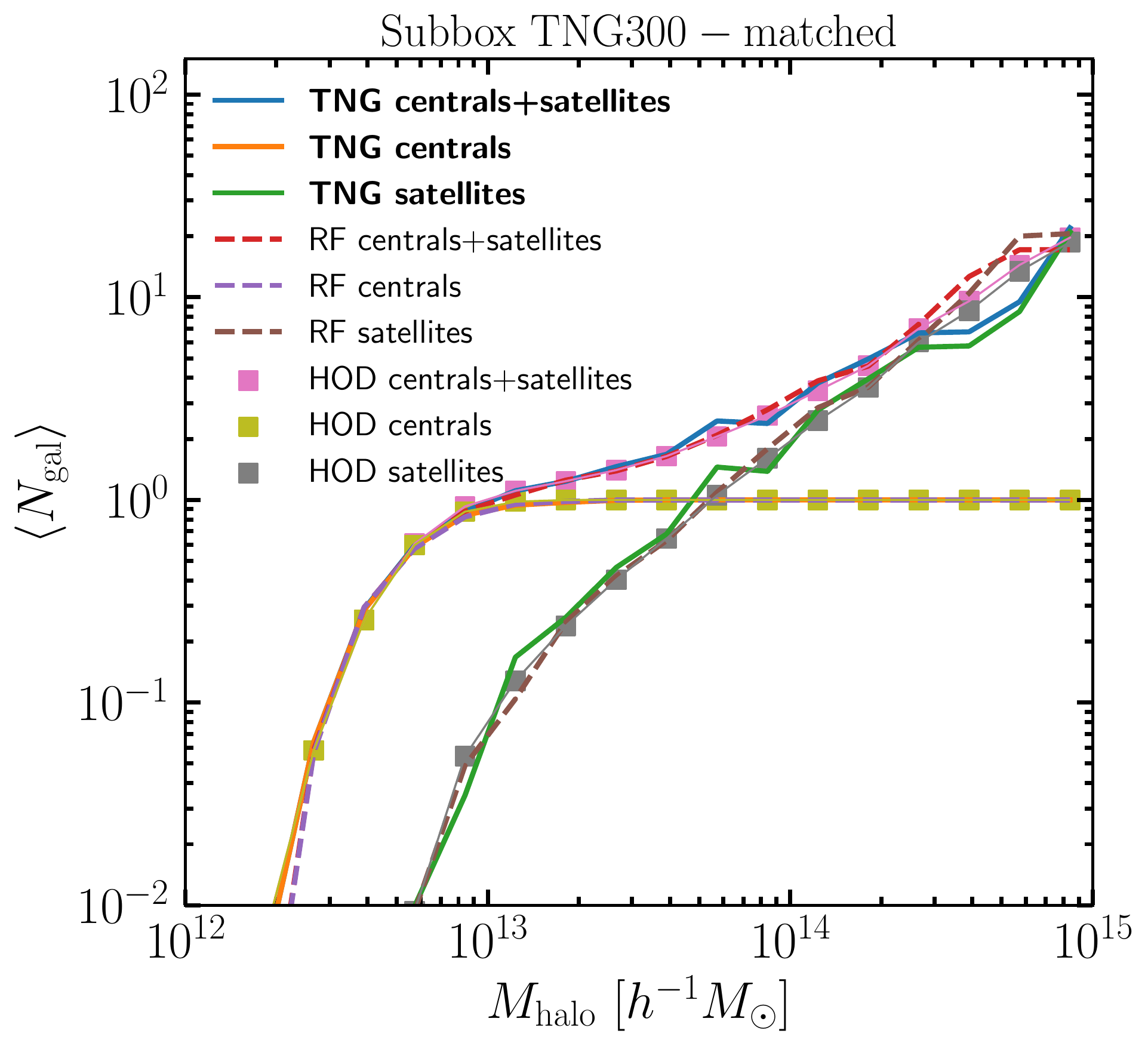}
\caption{The mean galaxy occupation of TNG300 as a function of the halo mass. The solid lines are the true occupations of TNG300, the dashed lines are the predicted occupations by the RF using halo mass as a training feature, and the dotted lines are the predicted occupations using the standard HOD mass only model. As a consistency check, we see that the RF is able to predict the mean galaxy occupation fairly accurately.}
\label{fig:hod_cent_sat}
\end{figure}

The standard HOD model posits that the number of galaxies residing in a halo depends solely on the mass of that halo. We construct the HOD using TNG300 populated with bijectively matched halos as described in section \ref{Method:TNG}. In order to emulate the kinds of galaxy samples that will be detected by surveys like DESI, we consider luminous red galaxies (LRG-like), which are stellar-mass selected, and emission-line galaxies (ELG-like), which are based on colour cuts and chosen as described in \cite{HadTac20}. However, we note that it has been shown that clustering in ELGs is not sensitive to secondary halo parameters. For this reason, we dedicate our discussion primarily to the LRG-like sample, and show corresponding results from the ELG-like sample in Appendix \ref{appendix:ELGs}. A more comprehensive study of ELGs will be the subject of a future investigation.

Our LRG-like galaxies are defined as subhalos with at least $10^4$ gravitationally bound star particles 
corresponding to a stellar mass of $M_* \approx5\times10^{10} [h^{-1}M_{\odot}]$ and we calculate the HOD for halos with dark matter mass greater than $10^{11} h^{-1}M_{\odot}$, corresponding to a number density $n_{gal}\approx1.4\times10^{-3}[h^3Mpc^{-3}]$ for both z=0 and z=0.8 samples.

We then fit a 5-parameter HOD model, splitting the mean occupation per halo mass into contributions from centrals, $N_{\rm cen}$, and satellites, $N_{\rm sat}$ as described in \cite{Zheng_2005}.

\begin{subequations}\label{eq:HOD}
\begin{align}
N^{\rm HOD}_{\rm cen} (M_h) \equiv& \frac{1}{2} \bigg[1+\textup{erf} \bigg(\frac{\log M_h -\log M_{\rm min}}{\sigma_{\textup{log} M}} \bigg) \bigg] \label{eq:HODcen}\\
N^{\rm HOD}_{\rm sat} (M_h) \equiv& \bigg(\frac{ M_h - M_{\rm cut}}{M_1} \bigg)^\alpha \label{eq:HODsat}
\end{align}
\end{subequations}

Here, $M_h = M^{\rm DMO}_{\rm 200m}$ is the total mass enclosed by a sphere with mean density 200 times the background density of the Universe, $M_\mathrm{min}$ is the characteristic minimum mass of halos that host central galaxies, $\sigma_{\textup{log} M}$ is the width of this transition, $M_{cut}$ is the characteristic cut-off scale for hosting satellites, $M_1$ is a normalization factor, and $\alpha$ is the power-law slope. The best-fit parameters for \{$\log M_{\rm min}, \sigma_{\textup{log} M},\log M_{\rm cut},\log M_1,\alpha$\} are: \{12.64,0.26,12.65,13.65,1.03\} for $z=0.0$ and \{12.48,0.26,12.54,13.30,0.95\} for $z=0.8$. 

It is important to note that the HOD parameters have \emph{not been tuned to match the summary statistics} like the power spectrum (as is typically done in the case of galaxy survey data analysis), rather the standard HOD is fitted to the halo occupation of the TNG300 data according to halo mass. Our goal here is to test the standard HOD model and improve its accuracy by augmenting with secondary parameters.

Fig. \ref{fig:hod_cent_sat} shows the HODs (for centrals, satellites, and combined) constructed using TNG300. We note that this figure is only for 30\% of the TNG300 volume, corresponding to the test-set for our machine learning techniques (see section \ref{Method:ML}).

\subsection{Machine Learning models}
\label{Method:ML}
A supervised machine learning algorithm (ML) trains a model by providing it a subset of input variables (henceforth called `features') and output variables (henceforth called `target'). The algorithm uses this subset as a training set to learn the relation between the features and the targets. The trained model is then used to predict the targets for a different subset (test set) of features.

In this work, we use two machine learning algorithms to achieve the aforementioned goals.  We use the random forest regressor to identify secondary halo properties that affect the galaxy-halo connection and we use symbolic regression to augment the standard HOD model with simple equations that incorporate the secondary halo parameters identified by the random forest. Both of these algorithms are described below.  

\subsubsection{Random Forest}
We use the random forest algorithm from the publicly available package \texttt{Scikit-Learn} \citep{Scikit}.
A random forest regressor (RF) is a collection of decision trees; each tree is in itself a regression model and is trained on a different random subset of the training data \citep{Bre01_RF}. The output from a RF is the mean of the predictions from the individual trees (a single decision tree is prone to over-fitting and therefore the ensemble mean of the different trees is used).
RFs have been used for applications to various cosmological problems \citep{LucPei18, MosNaa20, NadMao18,CohBat20,Muc20}

This method has a few key advantages over other ML models: 1) little hyper-parameter tuning is required, 2) it is computationally efficient and 3) its ensemble characteristic lessens over-fitting. 

We designate 30\% of the TNG300 volume as the test set and use the remaining 70\% of the volume as the training set, which is used to train the RF to predict $N_{\mathrm{gals}}$. Because there are no first principles as to which secondary halo properties best describe the galaxy-halo connection, we begin by training a mass only model, and in each subsequent model we implement an additional secondary halo property (halo properties described in section \ref{Method:halo_properties}). Essentially, the RF will approximate equation \ref{eq:intro}.

 We test the predictive power of our models on the clustering of the halos by computing the correlation function weighted by the $N_{\mathrm{gals}}$ predictions of the RF. We compare the different models to TNG300 and apply the properties from the best performing model for use with symbolic regression, described below.

\subsubsection{Symbolic Regression} 
Symbolic regression (SR) is a novel machine-learning technique that approximates the relation between an input and an output through analytic mathematical formulae \citep{SchLip09,UdrTeg20,WuTeg18,CraSan20,CraXu19,VilAngGen20,KimLu19,LiuTeg11}.
The advantage of using SR over other ML regression models like RF or deep neural networks is that it provides analytic expressions that can be readily generalized and that facilitate understanding the underlying physics. Furthermore, SR is shown to outperform other ML models when the size of data set is small \citep{Wil21}. In our case, we first use RF to get an indication of which parameters in the set of $\{i_h\}$ in Eq.~\ref{eq:intro} have the largest influence on $N_\textup{gal}$. We then compress the $\{i_h\}$ set to include only the most important parameters. Finally, as discussed in Sec.~\ref{sec:SyReg}, we use SR on the compressed set to obtain an explicit functional form to approximate $f$ from Eq.~\ref{eq:intro}.

In this study, we use the symbolic regressor based on genetic programming (an algorithm that searches equation space for a best fit) implemented in the publicly available \textsc{PySR} package\footnote{\label{PySR}\href{https://github.com/MilesCranmer/PySR}{
\textcolor{blue}{https://github.com/MilesCranmer/PySR}}} \citep{pysr,CraSan20}.

\subsection{Summary statistics for clustering}
\label{Methods:Summary_stats}

The goals of this work are 1.) to determine which secondary halo properties, in addition to halo mass, best model the galaxy-halo connection and 2.) augment the standard HOD model with simple equations that incorporate the effects of secondary halo properties.

As the goal of this study is to compare the predictive power of these augmented models on the clustering, we use the two-point correlation function in real- and Fourier-space as our summary statistics to determine the best model. 

\subsubsection{Correlation function}

We test the predictive power of our models on the clustering of the halos by computing the two-point correlation function weighted by $N_{\mathrm{gals}}$. For volumes with periodic boundary conditions, we use the natural estimator:

\begin{equation}
    \xi(r) = \frac{DD(r)}{RR(r)} -1
\end{equation}
where $DD$ is the number of halo pairs found at a separation radius, $r$, and $RR$ are the number of random points found at the same separation.

We note that for our work utilizing the RF, our training and testing volumes do not have periodic boundary conditions and we use the Landy-Szalay \citep{1993ApJ...412...64L} estimator as follows:

\begin{equation}
    \xi(r) = \left(\frac{N_{\mathrm{rand}}}{N_{\mathrm{data}}}\right)^2\frac{DD(r)}{RR(r)}-2\frac{N_{\mathrm{rand}}}{N_{\mathrm{data}}}\frac{DR(r)}{RR(r)} +1.
\end{equation}
where $N_{\mathrm{rand}}$ = 15 $N_{\mathrm{data}}$ random points and $DR(r)$ is the number of data-random pairs at separation $r$.

\subsubsection{Power spectrum}
The power spectrum $P(k)$ is the Fourier transform of the correlation function $\xi(r)$. 
We use the publicly available Pylians3 libraries\footnote{\label{Pylians}\url{https://github.com/franciscovillaescusa/Pylians3}} for calculating power spectra, wherein we first interpolate the galaxies onto a grid using cloud-in-cell (CIC) method and then use FFTs to find Fourier modes. We also subtract shot noise from the power spectrum using 
\beq
\begin{split}
P^\textup{SN}=\frac{\sum_i (N^i_\textup{gal})^2}{[\sum_i N^i_\textup{gal}]^2} V_\mathrm{box}
\end{split}
\eeq
where $N^i_\mathrm{gal}$ is the galaxy occupation of the $i^\mathrm{th}$ halo and $V_\mathrm{box}$ is the volume of the simulation box. As a complement to the results from $\xi(r)$, we will later show in Fig.~\ref{fig:SR_results} the $P(k)$ results for scales larger than those compared in the plots for $\xi(r)$.

\subsection{Secondary halo properties}
\label{Method:halo_properties}

TNG provides a model for the galaxy-halo connection. The standard HOD model, as stated above, is dependent solely on halo mass. However, work such as that by \cite{HadBosEis20} show that there is $\sim$15\% discrepancy in the two-point correlation function between standard HOD theory and TNG. One of the objectives of this work is to determine which secondary halo properties reduce the discrepancy. In this section, we outline the the various secondary halo properties considered in this study.

\begin{figure*}
\includegraphics[scale=0.43]{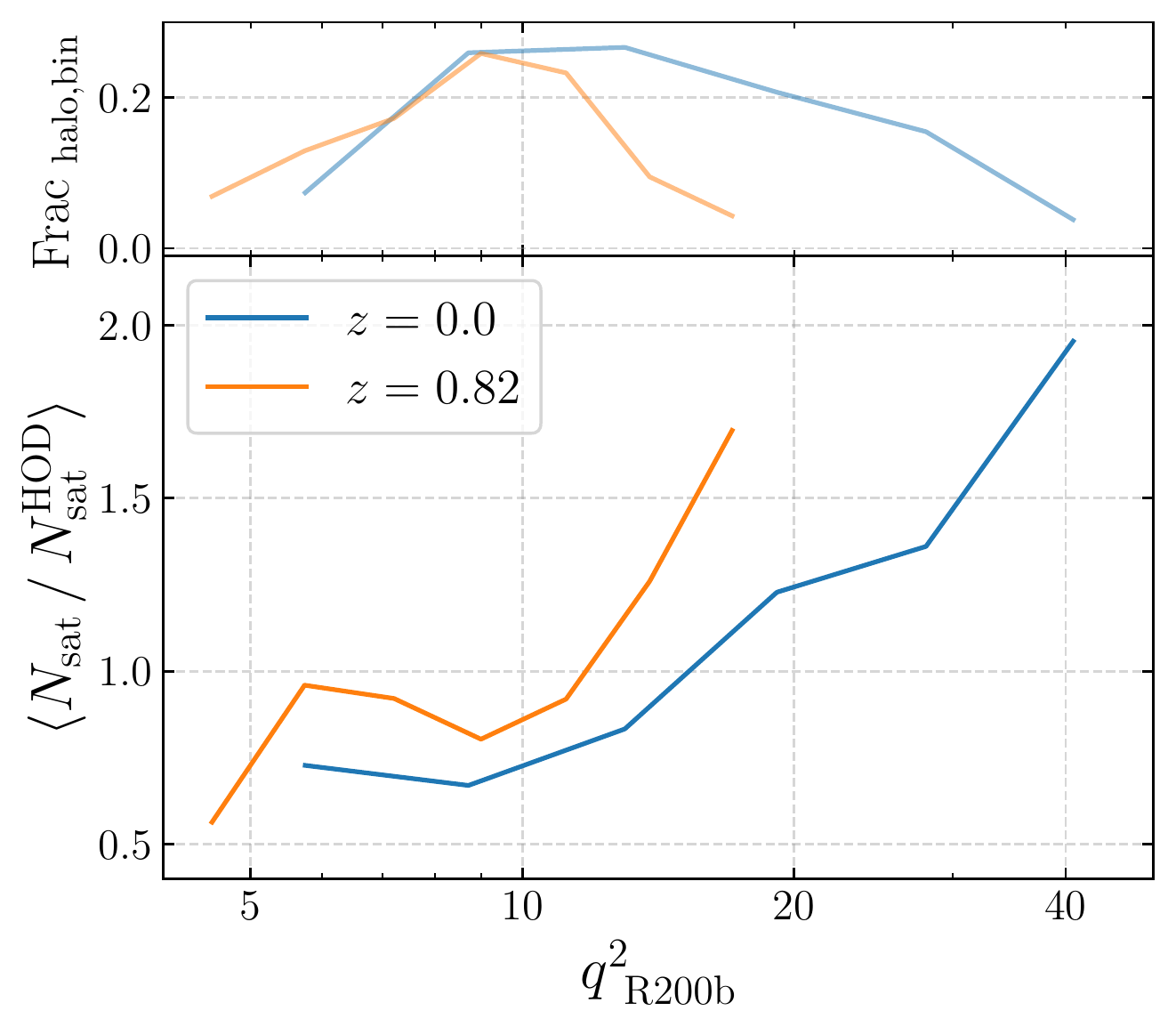}
\includegraphics[scale=0.43]{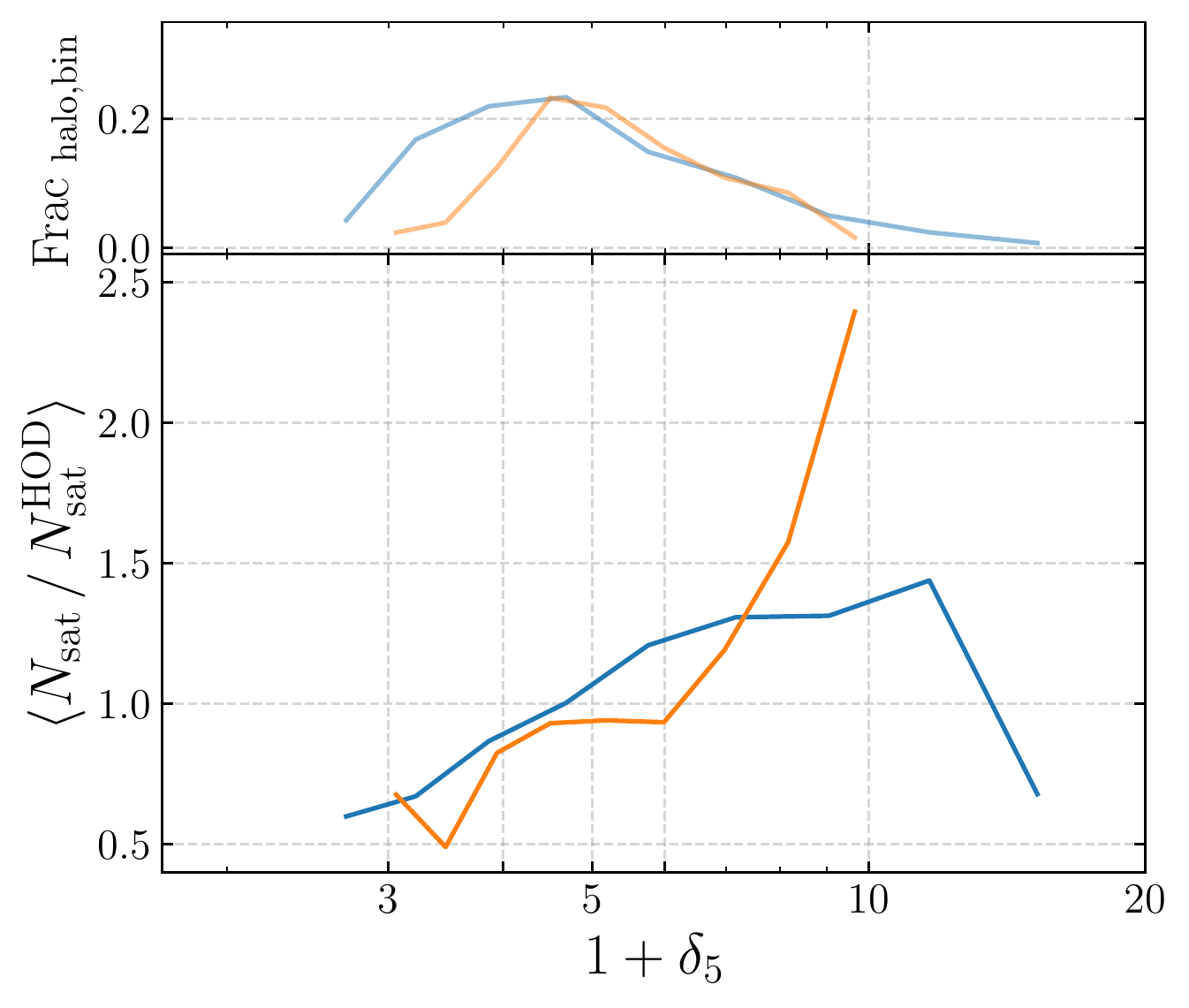}
\includegraphics[scale=0.43]{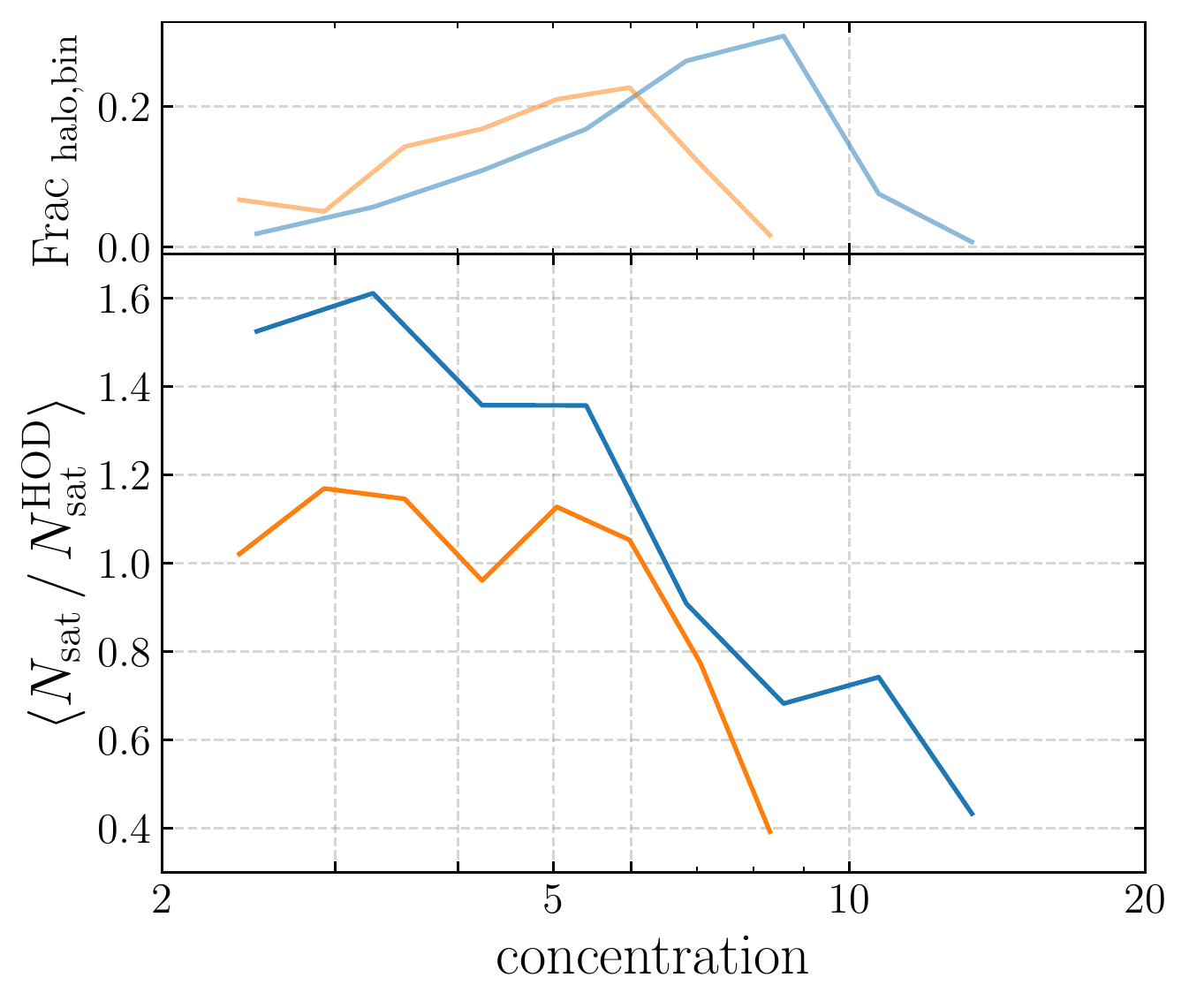}
\caption{Dependence of $N_\textup{satellites}$ on secondary halo properties: local shear (left), local environmental overdensity (center), and concentration (right) for halos in the mass bin $M_h \in [0.5-1]\times 10^{14}\Ms$. We see that satellites preferentially occupy anisotropic and denser environments. The top panels show the relative fraction corresponding to the number of halos in each halo property bin on the $x$-axis.}
\label{fig:Nsat_secondary}
\end{figure*}

\begin{figure*}
\includegraphics[scale=0.43]{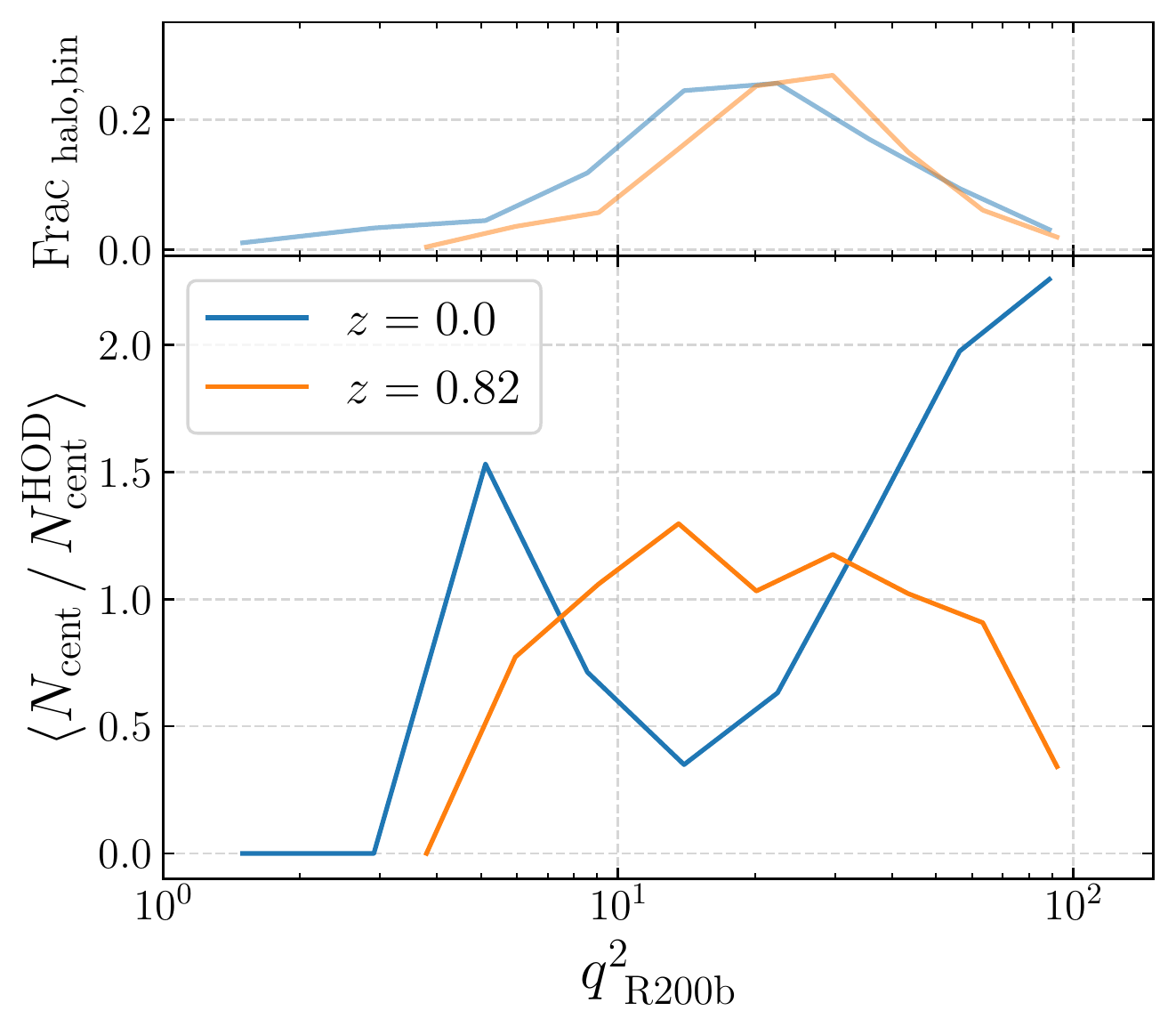}
\includegraphics[scale=0.43]{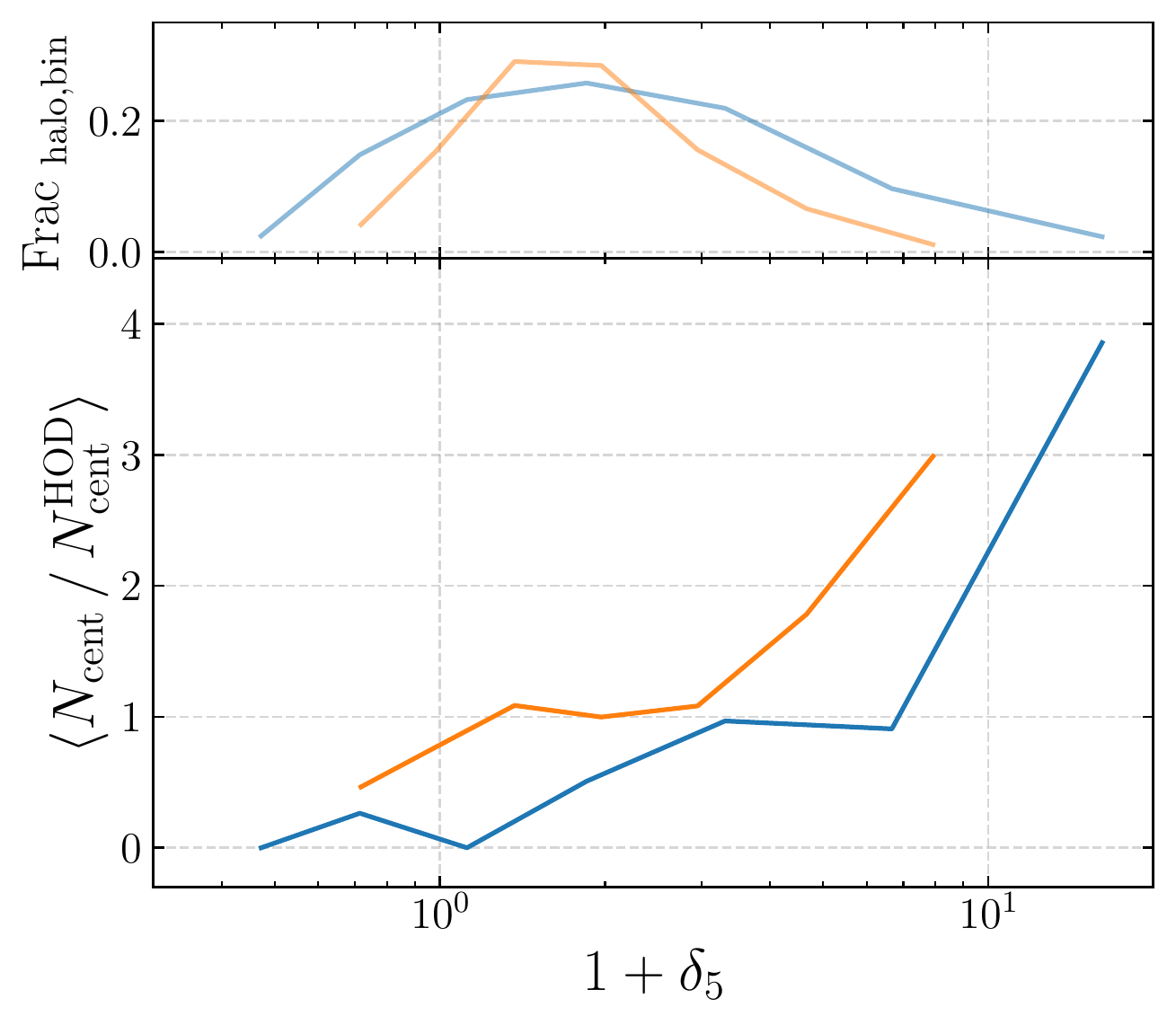}
\includegraphics[scale=0.43]{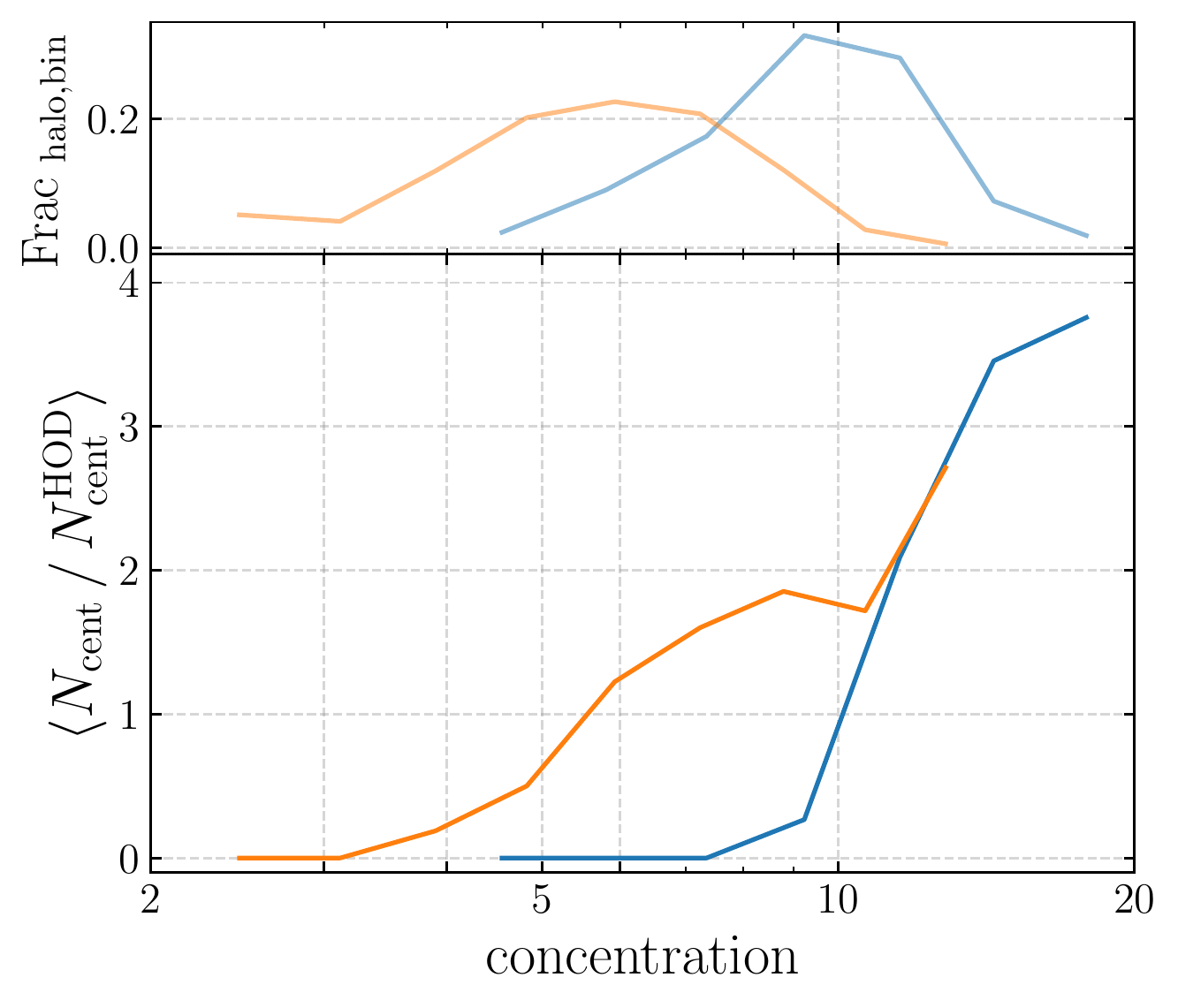}
\caption{Same as Fig.~\ref{fig:Nsat_secondary} but for centrals and for a bin corresponding to low halo masses (near the cut-off of mass required to host a central galaxy): $M_h \in [2-2.1]\times 10^{12}\Ms$. Overall, centrals in low-mass halos also occupy anisotropic and denser environments.}
\label{fig:Ncent_secondary}
\end{figure*}

\begin{figure*}
\centering
    \includegraphics[width=0.8\textwidth]{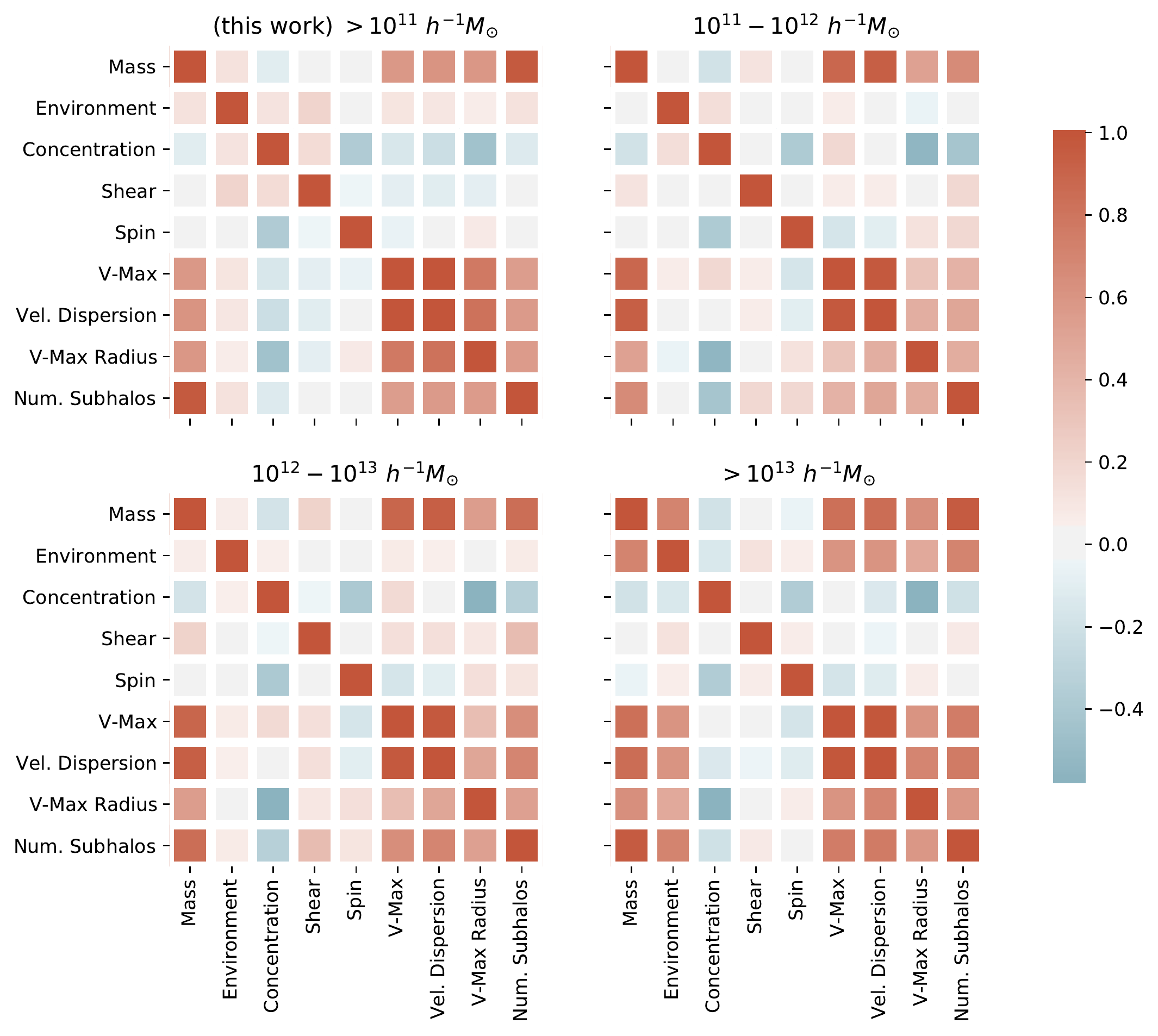}
    \caption{The correlations of several halo properties considered for use as features with which to train the RF. The four matrices show the correlations between each of these parameters for halos in different mass ranges. The top left matrix corresponds to the mass range of halos considered in this work. The color bar shows the strength of the correlation between any two parameters, where 1.0 is a perfect positive correlation. As we see in this diagram, the three parameters we consider in our final analysis (mass, environment and shear) have very little correlation with one another, allowing us to use these as independent parameters.}
    \label{fig:correlation_matrix}
\end{figure*}

(i) \textit{Local Environment} is the large-scale environment in which a halo is embedded. For each halo in the DMO simulation, we define the environment as $f_{env} \equiv \rho_R/\overline{\rho}$, where $\rho_R$ is the mass density in a sphere of radius $R$, and $\overline{\rho}$ is the mass density in the entire simulation volume. We furthermore use a tophat definition of environment with a smoothing radius at 5 Mpc from the center of the halo. 

The central panels of Fig. \ref{fig:Nsat_secondary} and Fig. \ref{fig:Ncent_secondary} examine the effect of environment on the number of satellite and central galaxies respectively. We see that both satellites and centrals preferentially occupy overdense regions. This may be due to overdense regions experiencing an increased number of mergers \citep{BosEis19,HadBosEis20}.

An alternate definition of local environment explored in this work, and discussed in Appendix \ref{apx:SyReg}, is that of the annulus environment. This definition is similar to that defined above, but limited to the density inside an annulus of $R_\mathrm{200m}$ to a radius of 5 $\mathrm{Mpc}/h$ surrounding the halo, where $R_\mathrm{200m}$ is the radius of a sphere whose density is 200 times the mean density of the Universe.

(ii) \textit{Environmental shear}. To quantify the anisotropy, we first calculate a dimensionless version of the tidal tensor as $T_{ij}\equiv \partial^2 \phi_R/\partial x_i \partial x_j$, where $\phi_R$ 
is the dimensionless smoothed potential field calculated using Poisson's equation: $\nabla^2 \phi_R = \delta_R$.$\ \rho_R$ is calculated by first interpolating the density field on to a grid and then smoothing it in Fourier space by a top-hat filter with radius $R$. Note that for calculating $\delta_R$ or $\phi_R$, we have not assumed spherical symmetry; we instead used the three dimensional particle distribution in the simulation snapshot. It is convenient to calculate the tidal tensor using the inverse fourier transform as $T_{ij} (\textbf{x}) = \mathrm{IFT} \left\{\delta_R (\textbf{k})  (k_ik_j)/k^2  \right\}$ \citep{ParHahShe18}.

We calculate the tidal shear $q^2_R$ using\footnote{It is worth noting that perturbation theory--based models use a closely related variable $s^2\equiv2q^2/3$ for studying the nonlocal bias (e.g. \cite{ChaSco12,BalSel12}).} \citep{CatThe96, HeaPea98}
\beq
q^2_R\equiv \frac{1}{2} \big[ (\lambda_2-\lambda_1)^2+(\lambda_3-\lambda_1)^2+(\lambda_3-\lambda_2)^2\big] \, ,
\eeq
where $\lambda_i$ are the eigenvalues of $T_{ij}$. Note that any spherically symmetric contribution to $q^2_R$ is automatically canceled, so $q^2_R$ is sensitive to the anisotropy of the field. We first calculate $q^2_R$ at multiple $R$ (e.g., 0.5 $\Mpc$, 1 $\Mpc$,...). We then interpolate over these values in order to calculate the shear at $R_\mathrm{200m}$ for consideration in this study.

The dependence of the number of satellites of a halo with its secondary properties is shown in Fig.~\ref{fig:Nsat_secondary}.
We find $N_\textup{sat}$ has a strong dependence on the tidal shear (see also Fig.~\ref{fig:shear_radius} for the dependence of $N_\textup{sat}$ on the radius at which the shear is calculated).

It is worth noting that some studies using SAMs have found an effect of the halo environment on satellites is weak (e.g. \cite{McEWei18}), while some others find a strong effect (e.g. \cite{XuZeh20,XuKum21}).

For the non-linear dark matter field (i.e at small scales), $q^2_R$ is strongly correlated with $\delta_R$ as inferred from Lagrangian perturbation theory (see \cite{ParHahShe18}). This is also seen in Fig.~\ref{fig:correlation_matrix}.

(iii) \textit{Concentration} characterises the density distribution of the halo. We use concentration calculated by fitting the NFW \citep{1996ApJ...462..563N, 1997ApJ...490..493N} profile to the halo using the phase space, temporal halo finder, ROCKSTAR \citep{2013ApJ...762..109B}.

(iv) \textit{Spin} is a measure of the angular momentum acquired by the halo. We adopt the following definition of dimensionless spin, $\lambda$, as in \cite{2001ApJ...555..240B}
\beq
\lambda = \frac{J_{\rm{vir}}}{\sqrt{2}M_{\rm{vir}}R_{\rm{vir}}V_{\rm{vir}}}
\eeq
where $J_{\rm{vir}}$ is the angular momentum inside a sphere of radius $R_{\rm{vir}}$ of mass $M_{\rm{vir}}$ and with halo circular velocity $V_{\rm{vir}}=\sqrt{2M_{\rm{vir}}/R_{\rm{vir}}}$.

(v) \textit{Vmax} is the maximum rotational velocity of a halo.

(vi) \textit{Velocity Dispersion} provides the one dimensional dispersion of dark matter particle velocities associated with the central galaxy of the halo.

Fig. \ref{fig:correlation_matrix} displays the correlations of several halo properties including those discussed above. Examining these correlations aids in choosing properties for use with ML. Halo properties that are only weakly correlated, or not correlated, allow for better interpretation of the RF results, as we see in the next section.

\section{Results}
\label{results}

We present an analysis of the summary statistics computed with predicted $N_{\mathrm{gals}}$ by ML algorithms as described in the methods section above.

\begin{figure}
    \centering
    \includegraphics[width=0.45\textwidth]{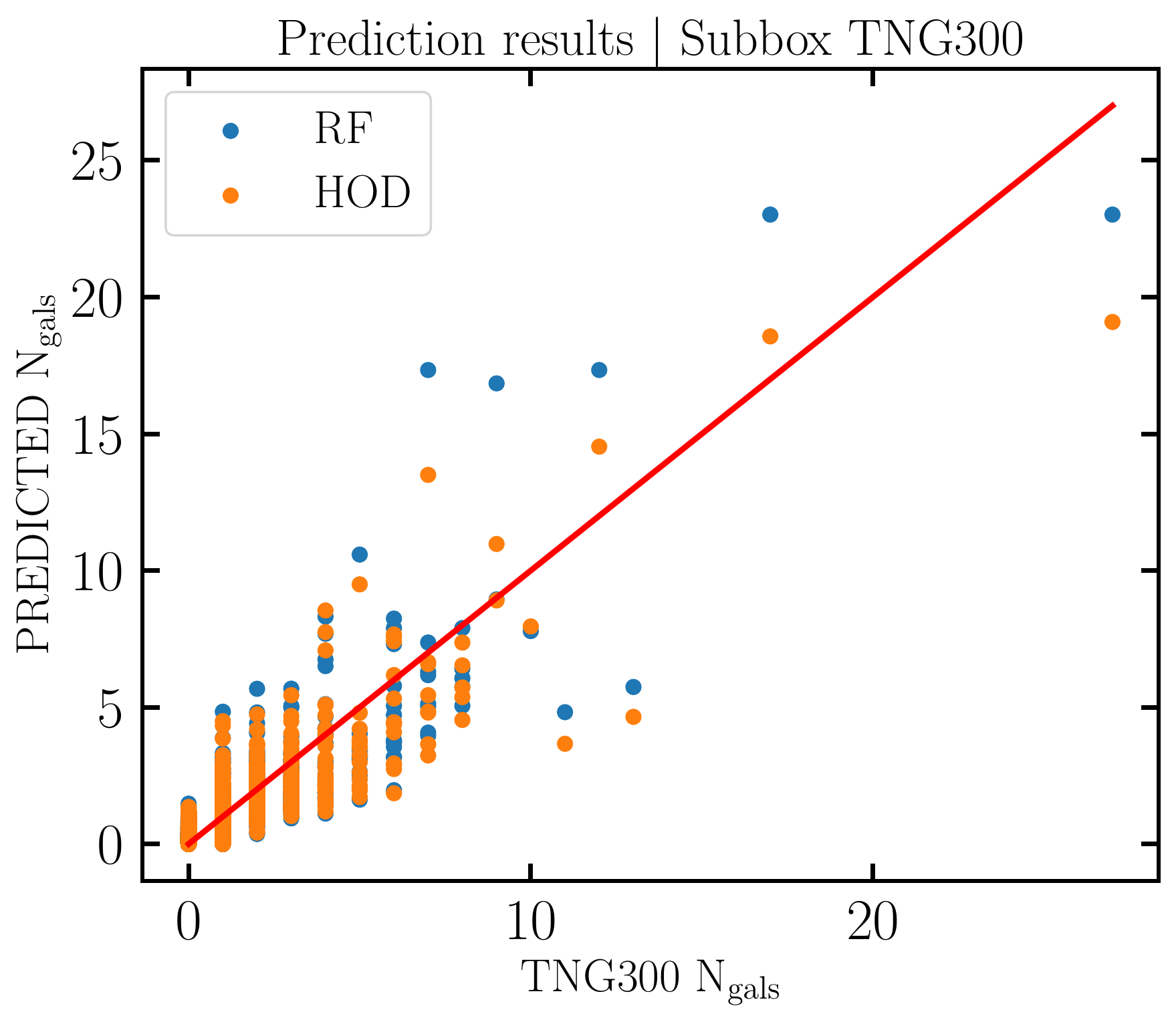}
    \caption{The predictive power of the Random forest (RF) (blue points) and the standard HOD model (orange points) as compared to TNG300. The x-axis shows the number of galaxies for halos in the TNG300 subbox used for testing the RF. The y-axis shows the predicted number of galaxies for the same subset. The solid red line indicates where the data points would lie in the case of a perfect prediction. We see that the RF performs similarly to the standard HOD.}
    \label{fig:RFvsHOD_performance}
\end{figure}
 
\begin{figure}
\centering
    \includegraphics[width=0.45\textwidth]{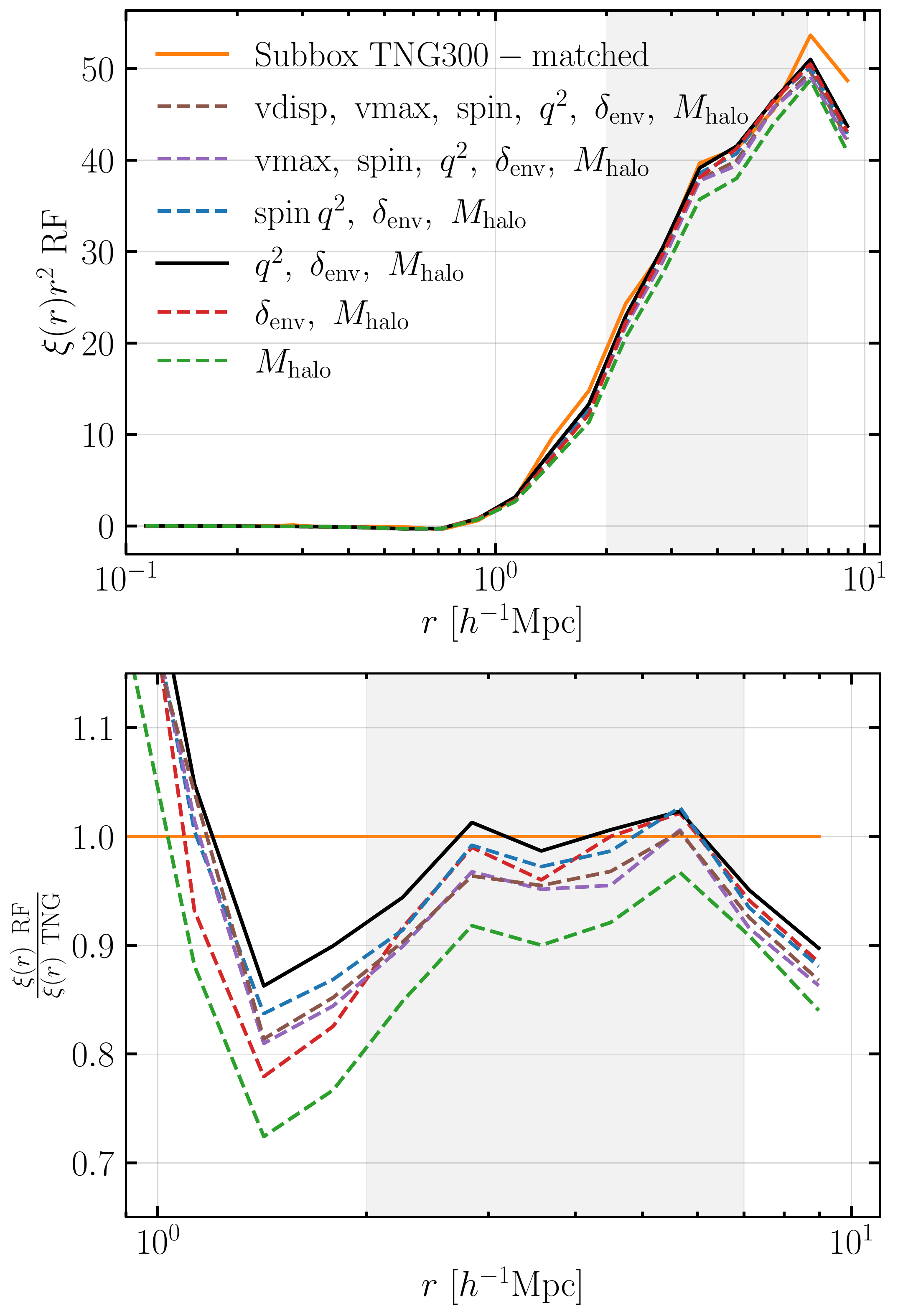}
    \caption{Random forest (RF) results. Using RF predictions of $N_{\mathrm{gals}}$ as weights to calculate the correlation function, we compare models incorporating different secondary halo properties (in color dashed lines) against TNG300 (solid orange line). \textbf{Top}: The correlation functions of the various models multiplied by $r^2$. \textbf{Bottom}: To increase clarity of results from the top panel, we show the ratio of correlation functions of the models to that of TNG300. Perfect agreement would fall along the solid orange line. We focus on a spatial range corresponding to the two-halo term, and note that the steep dip between (1-2) $h^{-1}$Mpc is due to the transition between the one-halo and two-halo terms. A reliable range for clustering is further limited due to a sample size of 30\% of the volume. We therefore highlight the reliable scales in the light grey shaded region. We are able to recover the standard, mass only, theory shown in the green dashed line, which shows a discrepancy from TNG300 comparable to what has been reported by the literature. We see that a model (black solid line) incorporating environment ($\delta_\mathrm{env}$ calculated with a smoothing scale of 5 Mpc) and shear ($q^2$ calculated at $R_\mathrm{200m}$) as secondary halo properties produces results closer to TNG300.}
    \label{fig:RF_corrfunc_compare}
\end{figure}

\begin{figure}
\centering
    \includegraphics[width=0.45\textwidth]{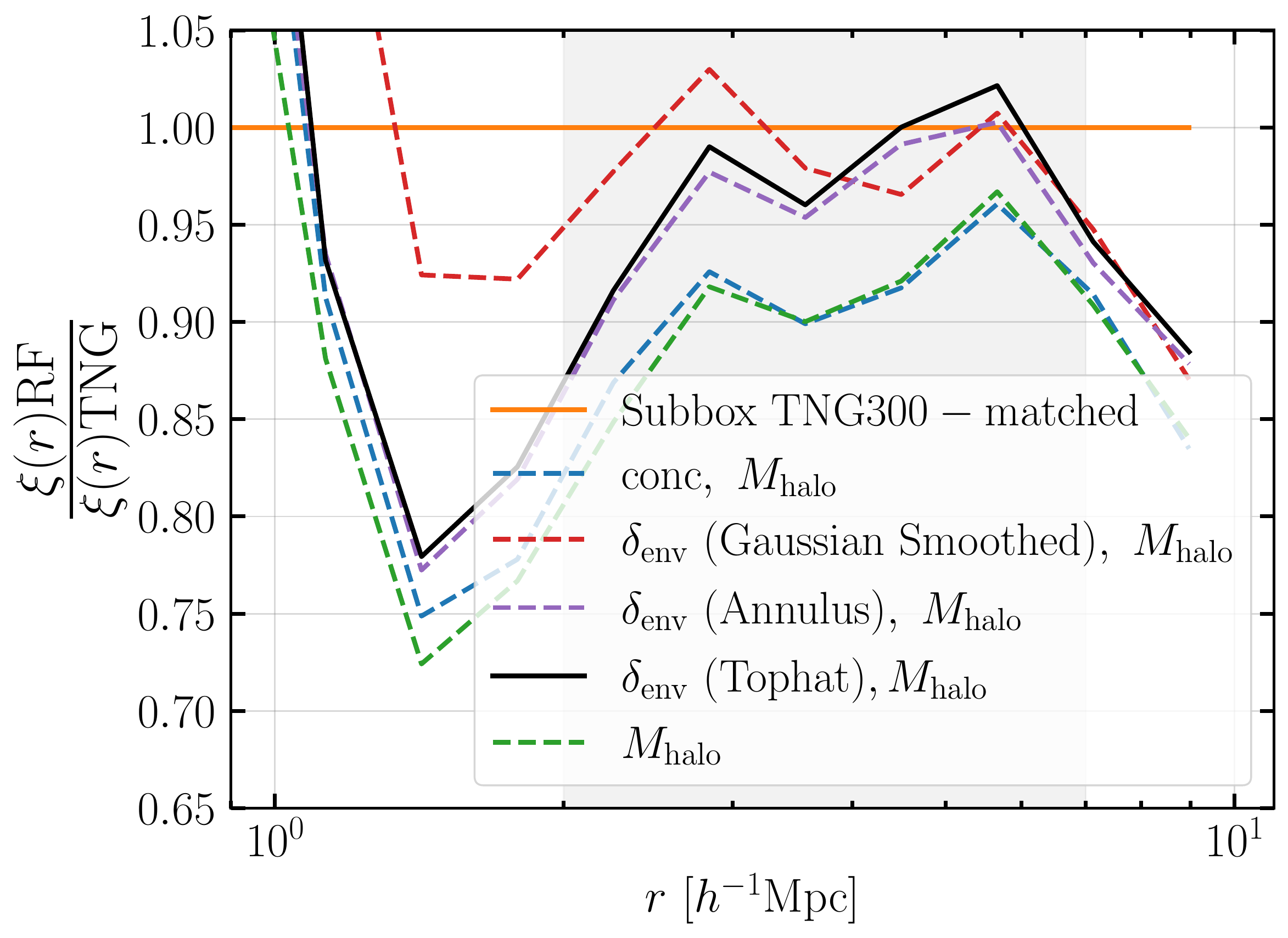}
    \caption{Random forest (RF) results. Similar to Fig. \ref{fig:RF_corrfunc_compare} but here we compare different definitions of environment, concentration and the standard mass only model. Concentration results in little improvement compared to the standard mass only model. While a Gaussian smoothed definition of environment (red dashed line) produced results closest to TNG300, we were unable to obtain useful results from the symbolic regression algorithm (discussed in text). We therefore use a top hat definition of environment (black solid line) for this work.}
    \label{fig:RF_corrfunc_env_compare}
\end{figure}

\begin{figure*}
\centering
\includegraphics[width=0.9\columnwidth]{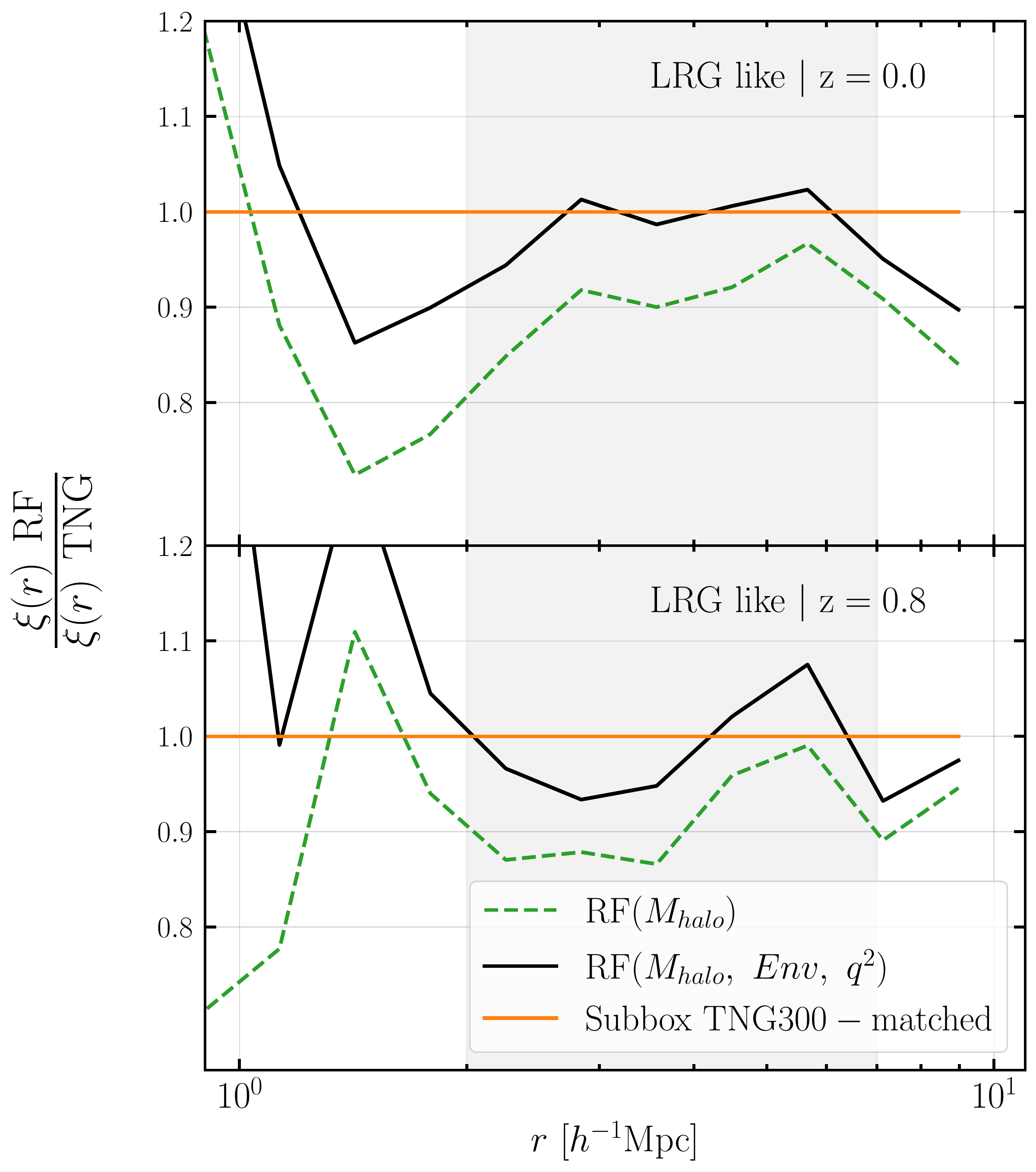}
\includegraphics[scale=0.55]{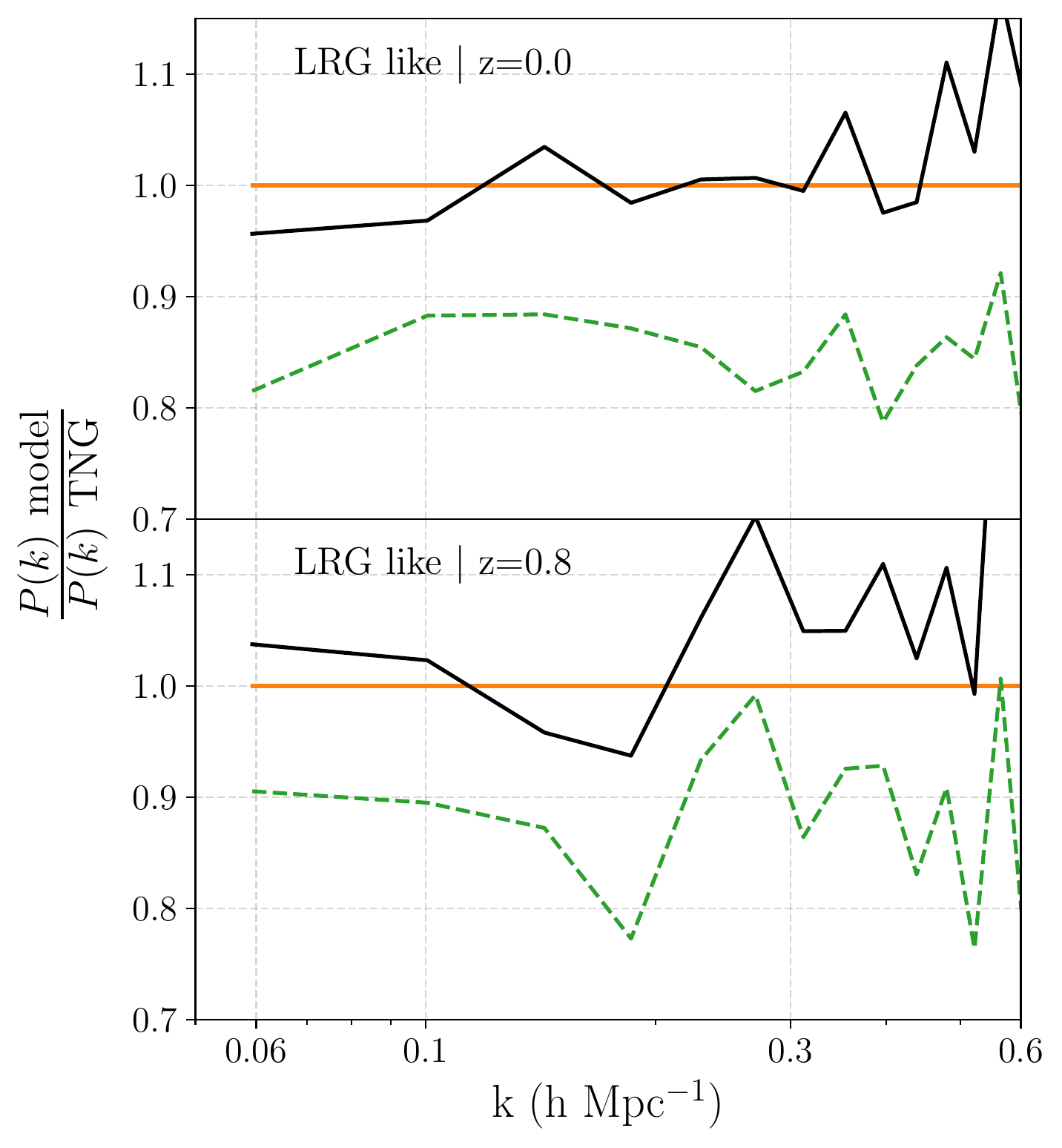}
\caption{Results from the random forest (RF). We compare the best resulting model as predicted by the RF, which incorporates mass, environment and shear, shown in the black solid line and a mass only model shown in the green dashed line. We show the clustering for these models at two different redshifts (z=0 and z=0.8). \textbf{Left:} Correlation function. The three-parameter model more closely matches TNG300 at z=0.0.  
\textbf{Right:} Power spectrum. We see similar results as the correlation function.}
\label{fig:RF_results}
\end{figure*}

\subsection{Random forest results}
\label{Results:RF}
As stated in section \ref{Methods:Summary_stats}, one of the goals of this work is to determine which secondary halo properties, in addition to halo mass, best model the galaxy-halo connection and reduce the discrepancy between the standard, mass only HOD and TNG. Because there are no first principles as to which secondary halo properties best achieve this goal, we utilize the RF to aid in determining these properties. We present the RF results in this subsection, emphasizing that these are using the RF test set, which makes up only 30\% of the TNG300 volume. 

The RF was able to recover the mass only HODs of TNG300, as seen in Fig. \ref{fig:hod_cent_sat}, as well as build HOD models incorporating the secondary properties discussed in section \ref{Method:halo_properties}. Fig. \ref{fig:RFvsHOD_performance} shows that the performance of the RF in predicting the number of galaxies for each halo in the test set is very similar to that of the standard HOD model. Fig. \ref{fig:RF_corrfunc_compare} shows the results of our various models constructed with the RF compared to TNG300. We see the $\sim$12\% discrepancy at z=0.0 between the mass only model (green dashed line) and TNG300 (orange solid line). The other dashed color lines show how we can improve upon our model by incorporating secondary halo parameters. The solid black line shows that using environment and shear in our model most closely matches TNG300, reducing the discrepancy by $\sim$9\%.

We note that for this study we are only focusing on the two-halo term. All of our satellite galaxies are therefore placed at the halo center. The effect this has on the correlation function can be seen in the top panel of Fig. \ref{fig:RF_corrfunc_compare} for distances less than $\sim$ 1 Mpc/h; the correlations in the regime of the one-halo term (multiplied by $r^2$ to emphasize effects at different scales) is $\sim$0. The bottom panel of the figure displays the ratios of the correlation functions of our various models against that of TNG300. The steep dip we see between 1-2 Mpc/h is the effect of transitioning from the regime of the one-halo term to the two-halo term. Here, the ratio of the two quantities is very noisy as both the numerator and the denominator are close to 0.

 Fig. \ref{fig:RF_corrfunc_env_compare} is the same as the bottom panel in Fig. \ref{fig:RF_corrfunc_compare} but specifically examines the difference between using environmental overdensity and concentration as features for the RF. Previous works have suggested that these two halo properties are influential in the galaxy-halo connection \citep{2018ApJ...853...84Z, 2018MNRAS.480.3978A,BosEis19}.
 We find that the RF does not estimate any statistically significant improvement to the mass only model by incorporating concentration, however there is improvement by incorporating environment. Furthermore, we examine three different definitions of environment (Gaussian smoothed, annular and tophat) and while the RF results suggest that a Gaussian smoothed definition of environment best matched TNG300, we were unable to converge on a reproducible result with symbolic regression for this definition of environment (discussed further in section \ref{sec:SyReg}). We therefore use a tophat definition of environment which gave the second best results, as indicated by the solid black line.  

Fig. \ref{fig:RF_results} shows the RF results using the best model, incorporating environment and shear in addition to halo mass, for LRG-like galaxies at two different redshifts. For consistency, the standard, mass only model is shown in a green dashed line, our model is in a solid black line and a perfect agreement between the RF and TNG300 is the solid orange line. The top panels show results at z=0 and the bottom panels at z=0.8. The RF is able to estimate the contribution of the secondary halo parameters to $N_{\mathrm{gals}}$ which improves clustering of the mass only model at distances greater than 1.5 Mpc/h by $\sim$9\% at z=0.0 and $\sim$7\% at z=0.8. 

\subsection{Symbolic regression results}
\label{sec:SyReg}

The second goal of this work is to augment the standard, mass only HOD with simple equations incorporating secondary halo properties. We chose to implement environmental overdensity and shear, as per the results of the RF, and used symbolic regression (SR) to obtain the augmented equations. We present the analysis of our SR results in this subsection, noting that we utilized the full volume of TNG300 for this final part of the study.

\begin{figure*}
\centering
\includegraphics[width=0.95\columnwidth]{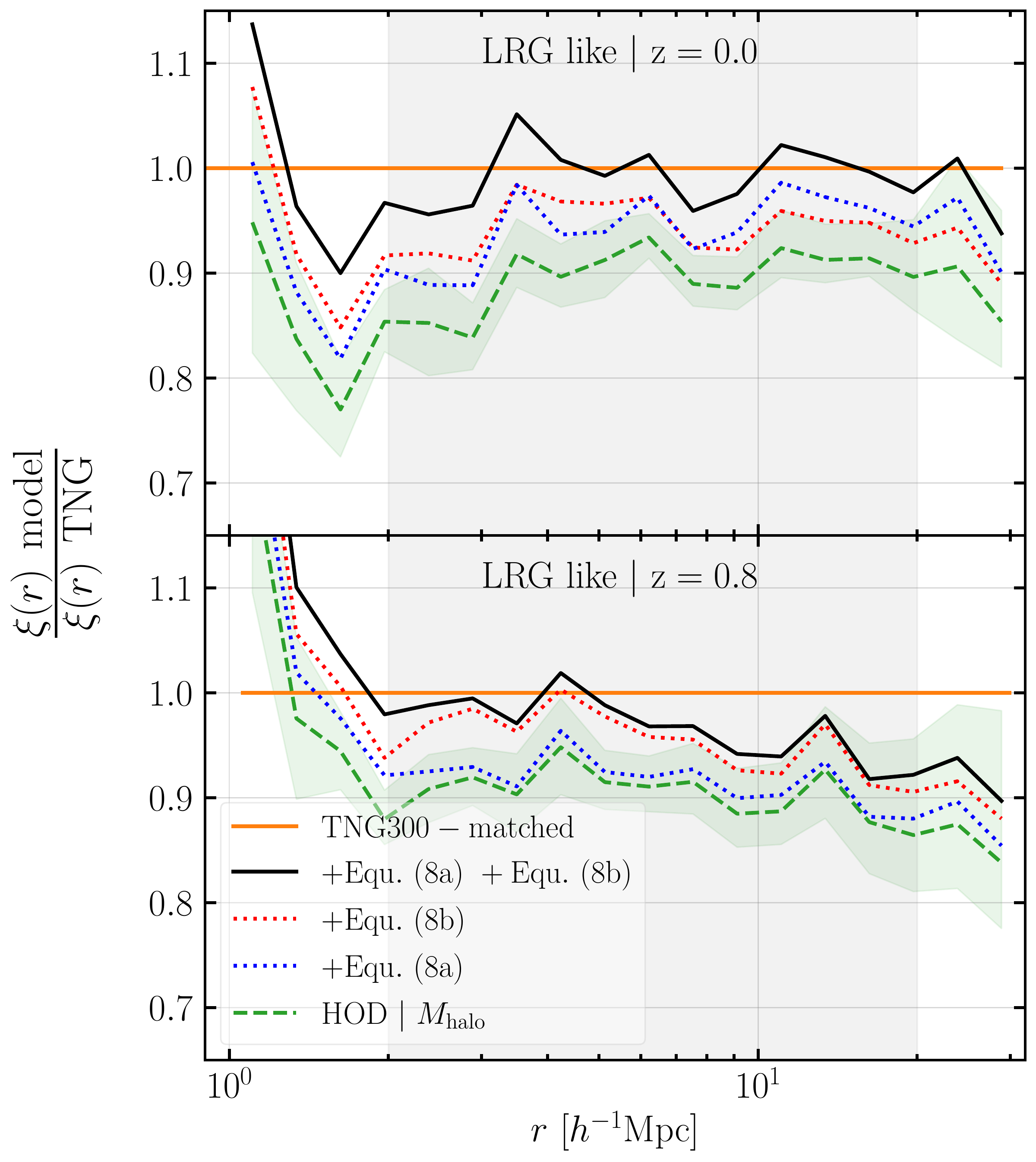}
\includegraphics[scale=0.5]{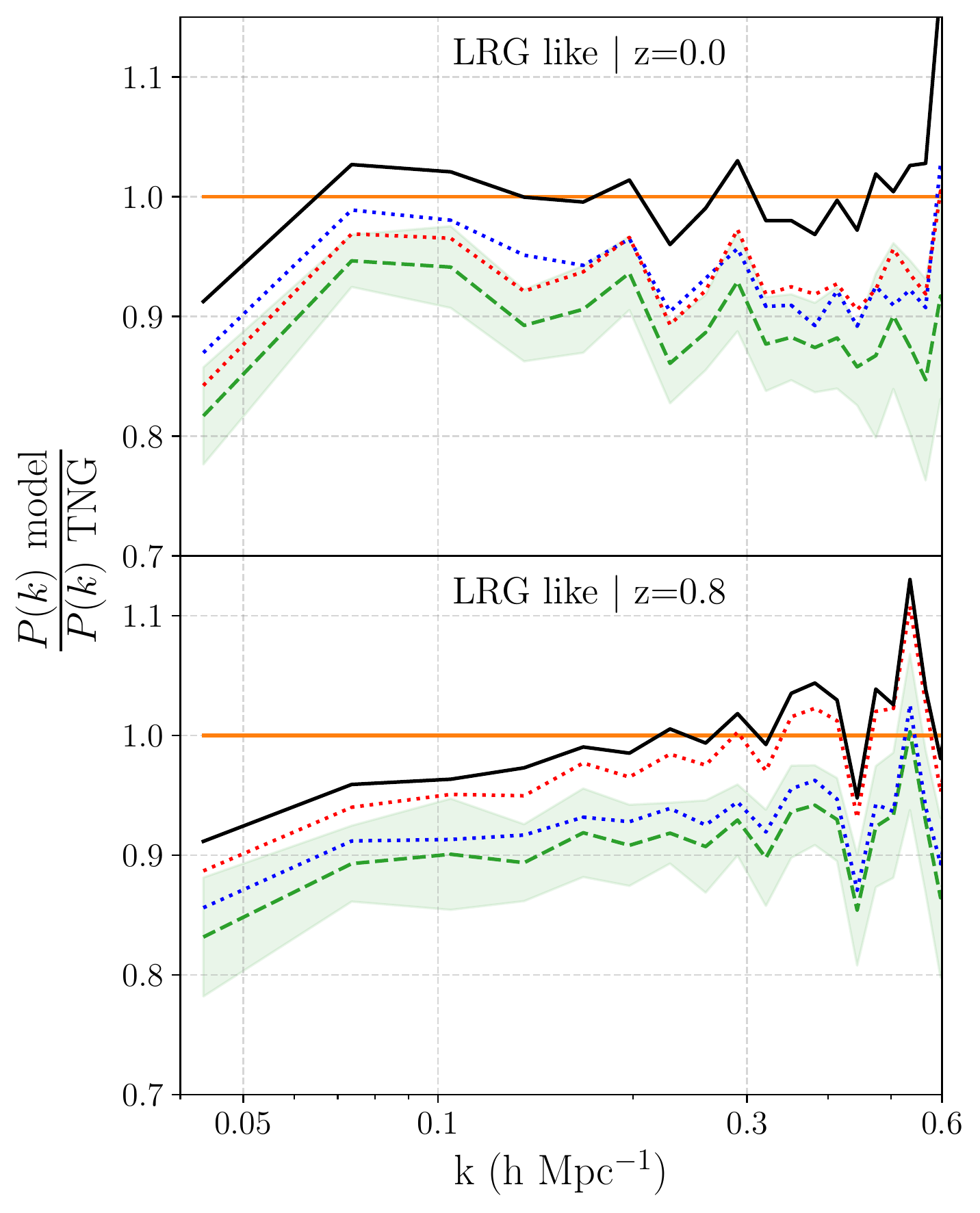}
\caption{The resulting correlation function (left) and power spectrum (right) using galaxy counts predicted from the augmented HOD equations as compared to TNG (orange line). The dashed green line shows the results using the mass only HOD as in equations (2) and (3). The solid blue line shows results with augmented HODSat, equ.(\ref{eq:NsatSR}), and mass only HODCent, equ.(2). The solid black line shows results utilizing augmented equations for both HODSat and HODCent. Equ.(\ref{eq:NsatSR}) refers to the augmented HODSat equation with a multiplicative term that incorporates environmental shear, $q'$, and Equ.(\ref{eq:NcenSR}) refers to that of HODCent with a multiplicative term incorporating the environmental overdensity, $\delta'$. Both of these multiplicative terms were obtained using symbolic regression. We see that the HOD model incorporating both environment and shear provides a substantial improvement over the mass only HOD. The grey shaded region shows the scales at which our clustering results are reliable.}
\label{fig:SR_results}
\end{figure*}

Before we input the various environmental parameters into the symbolic regressor, we re-scale the parameters by using logarithms to shorten the range over which these parameters vary: $q'\equiv \log_{10}(1+q^2_\mathrm{R200m})$ and $\delta'_\mathrm{env}\equiv \log_{10}(1+\delta_\mathrm{5})$ and find


\begin{subequations}\label{eq:SR}
\begin{align}
N_{\rm sat} (M_h) =&\,  N^{\rm HOD}_{\rm sat} (M_h) \times (q'-A) \label{eq:NsatSR}\\
N_{\rm cen} (M_h) =& N^{\rm HOD}_{\rm cen} (M_h) \times \Big[1+B(\delta'_\mathrm{env}-\overline{\delta'_\mathrm{env}})(1-N^{\rm HOD}_{\rm cen})\Big]
\label{eq:NcenSR}
\end{align}
\end{subequations}

$A$ and $B$ are constants which are fitted simultaneously with the other HOD parameters. The value of the constant $A$ is fixed such that the total number of satellites in the sample is roughly unchanged. The value of $B$ is similarly fixed, as well as to keep the minimum and maximum number of centrals 0 and 1 respectively. The results are shown in Fig.~\ref{fig:SR_results}. The slight discrepancy in the power spectrum at low-$k$ could be due to finite volume of the TNG box which leads to very few low-$k$ modes being sampled. 

In order to estimate the error bars for the summary statistics, we sample multiple realizations of the galaxy occupation. We use the mean galaxy occupation given by the HOD model and assume the centrals follow a Bernoulli distribution and the satellites follow a Poisson distribution.  We only show the error bars for the standard HOD case as the errors for the symbolic regression case are similar. 
Note that there is no error contribution from cosmic variance because all the cases are evolved from the same initial conditions. 

It is important to note that the expressions in Eq.~\ref{eq:NsatSR} are not unique. We have also found expressions which fit the TNG data better, however their form is relatively much more complex and are therefore prone to over-fitting. We show the results for these alternative expressions in Appendix \ref{apx:SyReg}. Furthermore, there is a large degeneracy between the environmental and shear parameters.

\subsubsection{AICc comparison}
\begin{table}
    \centering
    \def\arraystretch{1.5}
    \begin{tabular}{|p{2cm}||p{2cm}||p{1cm}||p{1cm}|}
    \hline
    \hline
    \multicolumn{4}{|c|}{AICc Scores} \\
    \hline
    \hline
    MODEL used & No. & SCORE & SCORE\\
    to weight $\xi(r)$ & PARAMETERS & z=0.0 & z=0.8\\
    \hline
    \hline
    $\mathrm{HOD}$: & 5 & 20.0 & 14.4\\
    $M_{halo}$ & & &\\
    \hline
    $\mathrm{+Equ.\ref{eq:NsatSR}}$:  & 6 & 9.0 & 13.4\\
    $M_{halo}, q^2$ & & &\\
    \hline
    $\mathrm{+Equ.\ref{eq:NcenSR}}$:  & 6 & -0.1 & -9.4\\
    $M_{halo}, \delta_{env}$ & & &\\
    \hline
    $\mathrm{+Equ.\ref{eq:NsatSR} +\ref{eq:NcenSR}}$: & 7 & \textbf{-7.2} & \textbf{-12.2}\\
    $M_{halo},q^2, \delta_{env}$ & & &\\
    \hline
    \end{tabular}
    \caption{The AIC scores corrected for small sample size (AICc) for the correlation function weighted by our models as compared to the correlation function of TNG300. All models have, at minimum, the 5 parameters of the standard HOD as in equ. \ref{eq:HOD}. A 6 parameter model has also included either secondary property $q^2 r_{200m}$ or $\delta_{env}$, while the 7 parameter model includes both secondary properties as parameters in addition to the standard 5.}
    \label{table:AICc_scores}
\end{table}

To compare the performance of various models, we calculate the Akaike information criterion corrected for small sample size (AICc). The AICc is an estimator of prediction error and penalizes a model for increased complexity. In other words, the AICc will give a lower score to a "better" model.

Because we are using the two-point correlation function as a measure for the predictive power of our models, we perform the AICc on the correlation function weighted by the predicted counts of our HOD models, not on the HOD models themselves. We note that we calculated the AICc using clustering at the reliable spatial range indicated by the grey shaded region in Fig. \ref{fig:SR_results}, approximately (2.0-20.0)Mpc.

The AICc is given by:
\begin{equation}
    \mathrm{AICc} = 2p - 2\log \mathscr{L} + \frac{(2\times p\times(p+1))}{(n-p-1)}
\end{equation}
where $\mathscr{L}$ is the likelihood,
\begin{equation}
    \mathscr{L}=\prod_{i=1}^{n} \frac{1}{n}exp\Bigl\{\frac{(y_i - \hat{y_i})^2}{\sigma^2}\Bigr\}
\end{equation}
$p$ is the number of parameters, $n$ is the sample size, $y_i$ are the clustering values given by TNG, $\hat{y}_i$ are our model predicted values and $\sigma$ is the standard deviation of TNG jackknife errors.

Table~\ref{table:AICc_scores} shows the AICc scores for our various models. We see that there is a preference for the three halo-parameter model, which incorporates environment and shear. This is a significant preference over the standard, mass only HOD, but only a small preference over the two halo-parameter model incorporating environment. Considering redshift dependence, all secondary-halo property models are preferred over the mass only HOD at $z=0.0$, however at $z=0.8$ there is very little improvement when using the two halo-property model incorporating shear (seen in the bottom panels of Fig. \ref{fig:SR_results}) and thus is the least preferred model by the AICc.


\section{Discussion}
\label{discussion}
We have presented an investigation of the galaxy-halo connection whereby we used machine learning (ML) in conjunction with the 300 Mpc box of TNG \citep{2018MNRAS.475..624N} to explore the HOD. In this section we make comparisons to previous studies and discuss the limitations of our methods. 

\subsection{Comparison to previous studies}
\label{discussion_comparison}

We now compare to other works which use ML to model galaxy properties in dark matter halos.

A recent paper by \cite{XuKum21} (henceforth Xu2021) also models the galaxy occupation, in their case by using a semi-analytic model (SAM) to train a RF to predict $N_{\rm{centrals}}$ and $N_{\rm{satellites}}$ separately and populate an \textit{N}-body simulation (Millennium). The box size they use is larger than ours (500 $\Mpc$ instead of 205$\Mpc$), which allows for more statistics and also comparison of correlation function to larger scales than in this work. They are able to recover the clustering of the SAM and galaxy assembly bias to a high precision by incorporating internal properties and formation history information. We attribute most of the qualitative differences in our RF results, as compared to Xu2021, to the nature of the method in which we map galaxies onto our \textit{N}-Body simulation (TNG300); namely that we use its matched hydrodynamical simulation. While both SAMs and hydro-sims have been tuned to match observations, comparisons of the two methods in the literature have shown that there are discrepancies between them despite reasonable agreement on galaxy formation \citep{2016MNRAS.461.3457G, 2018MNRAS.474..492M}. Additionally, there are differences in the way we implement some of our halo properties, for example, our local environment and environmental shear as described in section \ref{Method:halo_properties}. Fig. \ref{fig:RF_scalecompare} compares our fiducial model to ones with the same secondary halo properties calculated at a range of smoothing scales. Environmental shear is found to be critical term in TNG300, we therefore further look at the efficacy of a shear model augmented with a sample of secondary properties shown to be effective in Xu2021 in Fig. \ref{fig:Xu_compare}. In both instances, we see that there is a reasonable agreement between our fiducial model and other models for TNG300 at the reliable scales. We plan to revisit this work with a larger volume hydrodynamical simulation, with which we will obtain less noise and be able to do a more accurate measurement of clustering from our RF results. Lastly, that we do not use formation history as a training feature in our work is another difference worth mentioning. A previous study by \cite{HadBosEis20} showed that formation history is a weak assembly bias candidate in TNG300, we therefore omitted formation history as a parameter.

Furthermore, works such as \cite{Aga18} use ML to populate baryonic galaxies inside dark matter halos. This work also uses halo properties from cosmological hydrodynamical simulations as training features, such as mass, environment and spin, but differs in that it additionally incorporates properties such as growth history into their training features. Their ML algorithms were able to predict the mean baryonic properties, such as stellar mass, star formation rate, metallicity and neutral and molecular hydrogen masses, while we focus on accurately predicting the clustering of galaxies. 

Most notably, the main difference between our work and the aforementioned works is that our goal is to extend what traditional ML algorithms can do, and augment the HOD model itself with equations describing the occupation dependence on secondary halo parameters (for which we used symbolic regression). Our method is similar to that of \cite{WadVil20b}, but we focus on modeling galaxies while they focus on modeling the neutral hydrogen content of the halo.

It is also worth mentioning other approaches to augment the mass-only HOD model. \cite{McEWei18, XuZeh20} add an additional dependence of secondary halo properties to the HOD equations in equation~\ref{eq:HOD} (e.g. $M_\mathrm{min}$, $M_1$) with the halo environment, and their results are comparable to ours.

\subsection{Limitations / Future Work}
\label{discussion_limitations}

We used SR to obtain simple equations that incorporated secondary halo parameters based on data from the TNG simulation. Whether or not our equations work with different sub-grid physics than used in TNG, however, needs to be tested. There were more complex equations produced by the SR algorithm than the terms we present in our equations, however we decided on using equations with a simple form that were reproducible in multiple trials of the SR training. In this way we avoid using models that were prone to over-fitting. Furthermore, the RF ability to handle  high dimensional arrays is greater than that of SR; the dimensionality of the input space of SR needs to be relatively small. While we provided the SR algorithm with inputs of $\sim$5 parameters, the simple, reproducible equations we obtained only depended on a single halo parameter. The RF on the other hand, could be trained with all of the properties discussed in section \ref{Method:halo_properties} simultaneously. 

We note that we have ignored satellite profiles (i.e. the one halo term) in the clustering by placing all satellites at the center of the halo. Our method of weighting the correlation function by the predicted number of galaxies from the HOD model, is therefore not ideal for the clustering of the satellite galaxies. Hence we only show scales at the two-halo term, which is relevant for the large scale clustering to be done by upcoming surveys. However, the one-halo term will also be relevant for upcoming redshift surveys, where redshift space distortion effects are sensitive to the phase space distribution of satellites inside the halo. We therefore emphasize the importance of expanding upon this work to include the one-halo term for future study.

Lastly, we note that our volume size is also a limitation in this study. We see in the left panel of Fig. \ref{fig:SR_results} that with our current volume (300 Mpc on a side) we can robustly obtain clustering out to $\sim$30 Mpc. Because future surveys will be able to measure clustering out to more than 100 Mpc, we hope to improve upon our results with future simulations which are planned at the Gpc$^3$ volume. We also hope that at these larger scales the complexity of the equations will diminish and we will obtain more generalized forms of the augmented HOD terms. 

\section{Conclusions}
\label{conclusions}

The standard Halo Occupation Distribution (HOD) model, which predicts the number of galaxies that reside in a halo, is dependent only on halo mass. However, previous studies, such as those by \cite{Croton2007,BosEis19, HadBosEis20,XuZeh20}, have shown that there is a discrepancy between the occupation predicted by the standard, mass only HOD model and that predicted by simulations such as IllustrisTNG (TNG). In this paper we have presented an investigation of the galaxy-halo connection whereby we used machine learning (ML) in conjunction with the 300 Mpc box of TNG \citep{2018MNRAS.475..624N} to explore the HOD. We do this for LRG-like (luminous red galaxies), which are selected by stellar-mass, at redshift z=0.0 and z=0.8.  

We used halo catalogues created by first matching halos between the dark matter only and corresponding full physics TNG boxes, then populating the dark matter only box with well resolved galaxies (TNG300), as described in \cite{HadBosEis20}. Our objectives in this study were to 1.) use a random forest regressor (RF) to identify secondary halo properties that best reduce the discrepancy between the standard, mass only HOD model and TNG300 and 2.) use symbolic regression (SR) to augment the standard HOD model with simple equations that incorporate the secondary halo properties. Some of the secondary halo properties considered in this study include the local environmental, environmental shear, spin and more as defined in section \ref{Method:halo_properties}. Our ML learning algorithms predicted the number of galaxies ($N_{\mathrm{gals}}$) for a given model, which we used as weights assigned to the TNG300$_{DM}$ halos to predict the clustering in the full physics box. We used the predicted clustering as the baseline statistic with which to assess the quality of each augmented HOD model. Our findings are summarized below: 

(i) As a consistency check, the RF is able to predict the mean number of centrals and satellites. We are thus able to recover the HODs for TNG300 as seen in Fig. \ref{fig:hod_cent_sat}.

(ii) Fig. \ref{fig:Nsat_secondary} shows that satellite galaxies in high mass halos preferentially occupy anisotropic and denser environments.

(iii) Fig. \ref{fig:Ncent_secondary} shows that central galaxies in low mass halos also preferentially occupy anisotropic and denser environments.

(iv) Fig. \ref{fig:correlation_matrix} reveals that the three parameters considered in our final analysis (mass, environment and shear) do not have a strong correlation with one another, and were therefore favorable for use as independent parameters. 

(v) The RF revealed that incorporating environmental overdensity and shear as training features, in addition to halo mass, resulted in improved prediction of $N_{\mathrm{gals}}$ than did using halo mass alone, resulting in an improvement of $\sim$10\% in the clustering (Fig. \ref{fig:RF_results}).

(vii) Symbolic regression (SR) is useful to build flexible HOD models which are motivated by hydrodynamical simulations (our main results are in Eqs.~(\ref{eq:NsatSR}) and (\ref{eq:NcenSR})). 

(viii) Incorporating environmental overdensity and shear into the HOD equations results in ~10\% improvement in the weighted correlation function and power spectrum compared to TNG300 at redshifts z=0.0 and z=0.8 (Fig. \ref{fig:SR_results}). 

(ix) The AICc favors a model incorporating three halo properties to a mass only model at both z=0.0 ($\Delta \mathrm{AICc}=27.2$) and z=0.8 ($\Delta \mathrm{AICc}=26.6$)

For future study, it would be interesting to test the dependence of our results on astrophysical feedback parameters. One possibility could be to use the CAMELS suite of simulations \citep{VilAngGen20} which contain 2000+ hydrodynamical simulations run for different astrophysical feedback and cosmology parameters. Another possibility could be to use semi-analytic models (SAMs), which are computationally inexpensive galaxy-formation models \citep{2008MNRAS.391..481S}.

\section*{Acknowledgements}
We thank the anonymous referee for enlightening feedback which improved the quality of this work. We thank Ravi Sheth, Uro{\v s} Seljak, Chirag Modi, Chang Hahn and Miles Cranmer for helpful discussions.
We have used the publicly available Pylians3$^{\ref{Pylians}}$ libraries to carry out the power spectrum measurements and \textsc{PySR}$^{\ref{PySR}}$ package for symbolic regression. DW acknowledges support from the Friends of the Institute for Advanced Study Membership. SB is supported by the UK Research and Innovation (UKRI) Future Leaders Fellowship [grant number MR/V023381/1].

\section*{Data Availability}

The IllustrisTNG data is publicly available at \url{www.tng-project.org}, while the scripts and halo property data used in this project are readily available upon request.


\bibliographystyle{mnras}
\bibliography{Main}



\appendix
\section{Alternate symbolic regression expressions}
In section \ref{discussion_limitations}, we discuss that the equations obtained using SR are dependent on the definition of the halo parameter, i.e. we would obtain slightly different forms of an augmented $N_{cent}$ expression for a top hat definition of local environment than for an annular definition of local environment. Furthermore, the PySR package outputs a selection of expressions ranked by complexity and loss score. In this appendix, we present some of those alternate SR expressions in equation \ref{eq:NsatSR2} and equation \ref{eq:NcentSR2} of this appendix.

\label{apx:SyReg}

\beq\begin{split}
N_{\rm sat} &=  N^{\rm HOD}_{\rm sat} (M_h) \times (q'^2/c_1)\\
N_{\rm sat} &=  N^{\rm HOD}_{\rm sat} (M_h) \times \frac{1}{c_2}\exp (q' + c_3)
\end{split}
\label{eq:NsatSR2}
\eeq

\beq\begin{split}
N_{\rm cent} &= N^{\rm HOD}_{\rm cent}(M_h) \times (1 + c_4 \delta' -c_5)\\
N_{\rm cent} &= N^{\rm HOD}_{\rm cent}(M_h) + (c_6 \delta' - c_7)
\end{split}
\label{eq:NcentSR2}
\eeq
where $c_i$ are constants whose value can be tuned to the particular sample being analyzed.

\begin{figure}
    \centering
    \includegraphics[width=0.9\columnwidth]{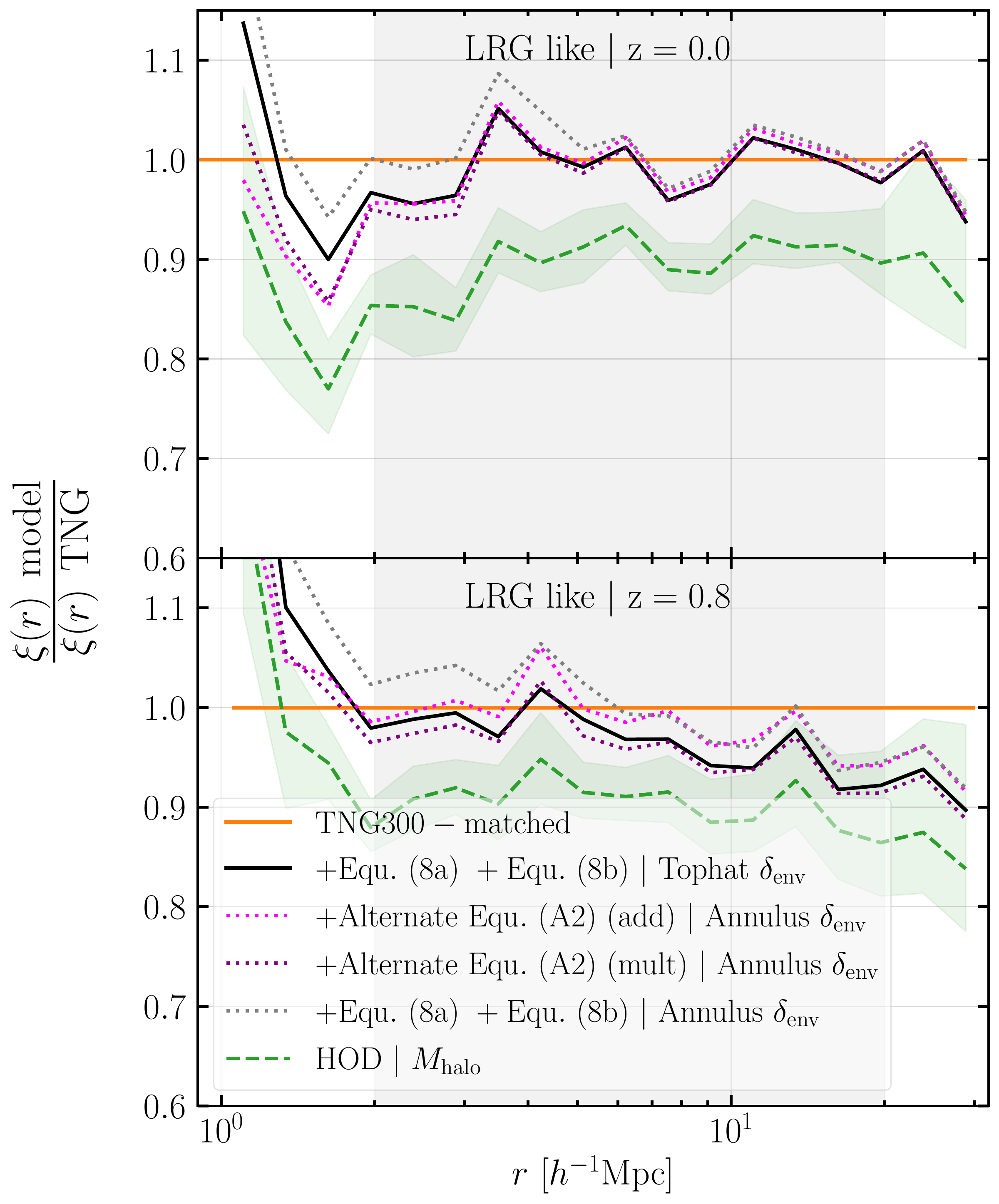}
    \caption{Difference in HODcent alternate equations obtained by symbolic regression (magenta and purple). Our fiducial augmented HOD equation with two different definitions of environmental overdensity (grey dotted and black solid lines).}
    \label{fig:alternate_SR}
\end{figure}

Figure \ref{fig:alternate_SR} compares results of our fiducial augmented HOD model (black solid line) to models incorporating either an alternate form of $N_{cent}$ or alternate definition of the local environment. Those shown in the magenta and purple dotted lines are alternate expressions obtained for an annulus environment. The grey dotted line is our fiducial augmented HOD but with annulus environment as input. We see that these alternate forms are generally consistent with our fiducial model, each bringing the clustering to closer agreement with TNG300 by $\sim$5\% -10\%. This is reinforced in Table \ref{table:AICc_alternate}, which shows the AICc scores for these alternate models are comparable to our fiducial model.

\begin{table}
    \centering
    \begin{tabular}{|p{3cm}||p{1cm}||p{1cm}||p{1cm}|}
    \hline
    \hline
    \multicolumn{4}{|c|}{AICc Scores for Alternate models} \\
    \hline
    \hline
    MODEL used & No. & SCORE & SCORE\\
    to weight $\xi(r)$: & PARAMS & z=0.0 & z=0.8 \\
    \hline
    \hline
    $\mathrm{HOD}$: & 5  & 20.0 & 14.4\\
    $M_{halo}$ & & &\\
    \hline
    $\mathrm{+Alternate\ Equ.\ref{eq:NcentSR2}(add)}$: & 7 & -5.04 & \textbf{-17.55}\\
    Annulus Env & & &\\
    \hline
    $\mathrm{+Alternate\ Equ.\ref{eq:NcentSR2}(mult)}$: &7 & -3.4 & -6.2\\
    Annulus Env & & &\\
    \hline
    $\mathrm{+Equ.\ref{eq:NsatSR} +\ref{eq:NcenSR}}$:& 7 & -2.2 & -11.7\\
    Annulus Env & & &\\
    \hline
    $\mathrm{+Equ.\ref{eq:NsatSR} +\ref{eq:NcenSR}}$: & 7 & \textbf{-7.2} & -12.2\\
    Tophat Env (fiducial)& & &\\
    \hline
    \end{tabular}
    \caption{same as Table \ref{table:AICc_scores} but for the alternate models.}
    \label{table:AICc_alternate}
\end{table}

\section{Sensitivity to parameter definition}
\label{app: diff_scales}
\begin{figure}
    \centering
    \includegraphics[width=0.9\columnwidth]{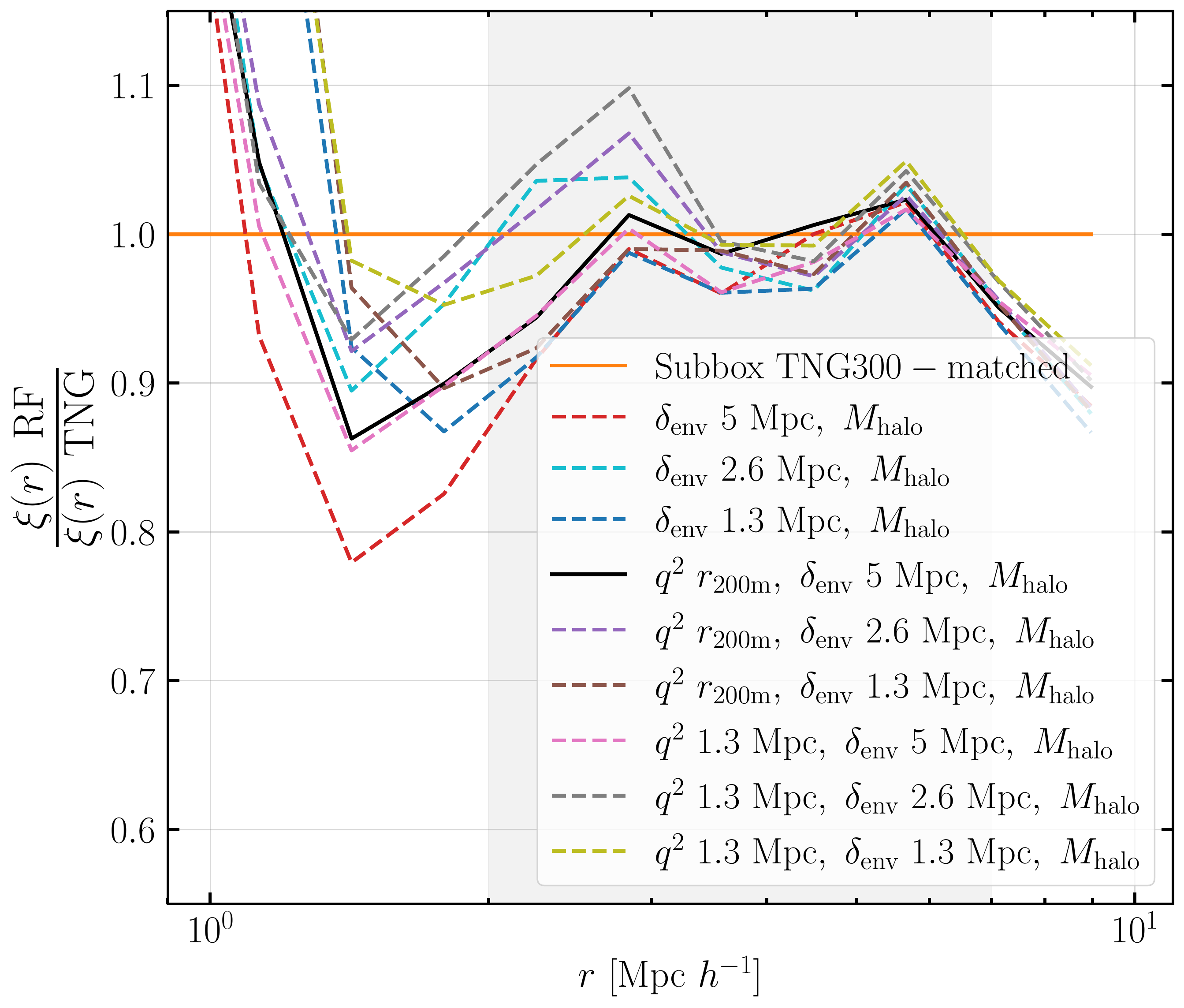}
    \caption{RF results for a variety of models, exploring environment and shear calculated at different scales. Our fiducial model is shown in a black, solid line and is comparable to other models.}
    \label{fig:RF_scalecompare}
\end{figure}

Fig. \ref{fig:RF_scalecompare} shows the RF results for models with same halo parameters as our fiducial model, but with the secondary parameters calculated at various scales. We calculate local environment, $\delta_{env}$, at three different smoothing scales (1.3, 2.6, 5.0) Mpc, and we calculate environmental shear, $q^2$, at a radius of 1.3 Mpc in addition to that calculated at $r_\mathrm{200m}$. We compare the different models in dashed, colored lines to our fiducial model shown in a black, solid line. We note several other models have reasonable agreement with our fiducial model. We will leave this for further investigation in a near future study where we will be able to obtain clustering statistics at a wider spatial range using larger volume hydrodynamical simulations.

\begin{figure}
    \centering
    \includegraphics[width=0.9\columnwidth]{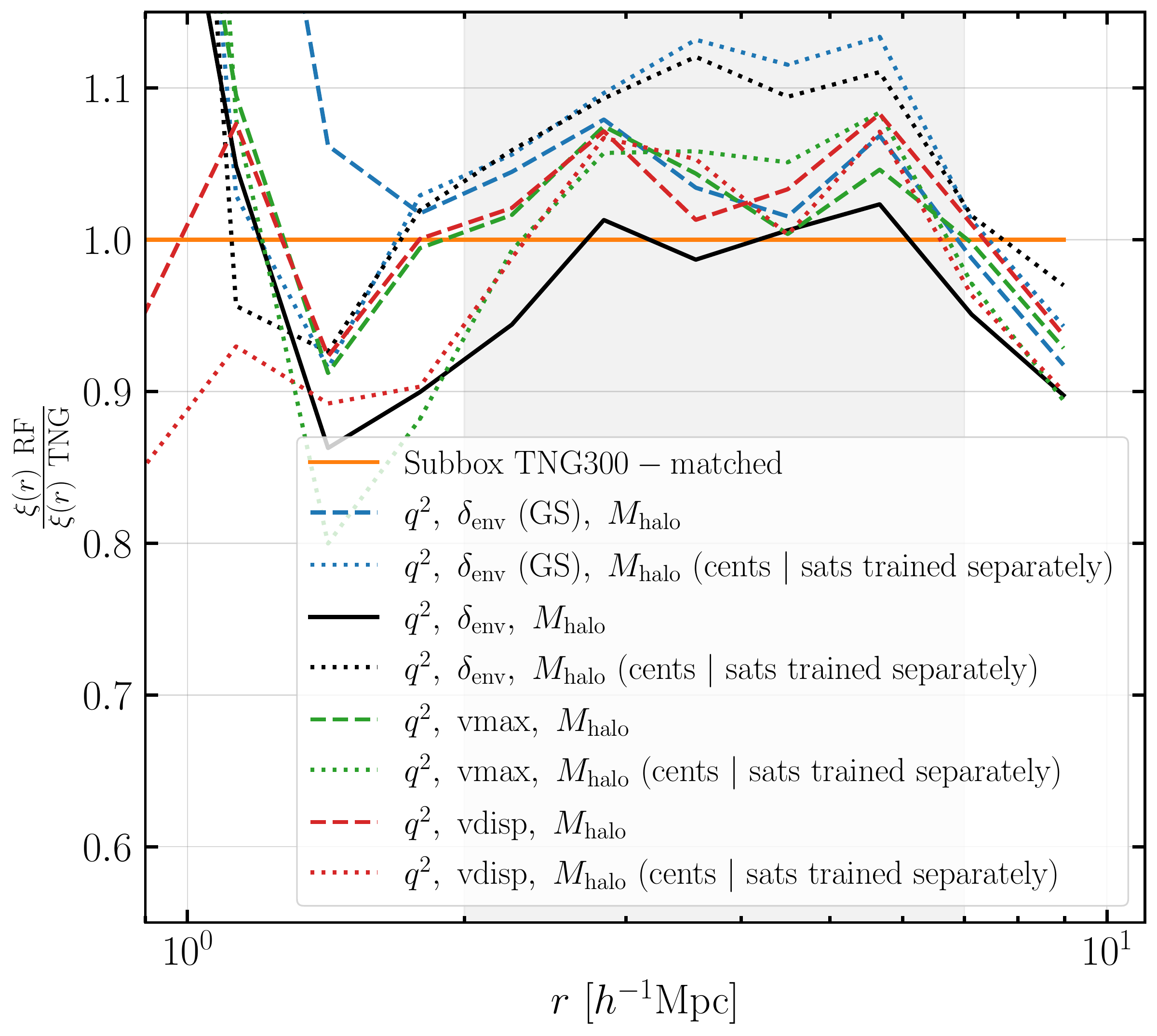}
    \caption{RF results comparing our fiducial model to other shear models augmented with some secondary parameters found effective in \citet{XuKum21}. Here, shear, $q^2$, is calculated at $r_{200m}$ and we include two definitions of local environment, our fiducial, $\delta_\mathrm{env}$ is a tophat definition as described in section \ref{Method:halo_properties}, and $\delta_\mathrm{env}\ \mathrm{(GS)}$ is a Gaussian smoothed definition. Dotted lines represents models where the RF was trained on central galaxies separately from satellite galaxies. Using TNG300 as ground truth, a simple $\chi^2$ measurement finds that a model in which the RF is trained with centrals and satellites together using $q^2$ and $\delta_\mathrm{env}$ as secondary halo parameters (solid black line) provides the closest clustering results to our target at the reliable scales (grey shaded region).}
    \label{fig:Xu_compare}
\end{figure}

Fig. \ref{fig:Xu_compare} explores the efficacy of shear models (shear is found to be a critical term in TNG300) augmented with a sample  of secondary parameters found effective in \cite{XuKum21}, namely Vmax and velocity dispersion. Our fiducial model is shown as a black solid line. Also included is a version of our fiducial model with a Gaussian smoothed (GS) definition of local environment. Dotted lines represent models in which the RF was trained on central galaxies and satellite galaxies separately. Dashed lines and the fiducial model show results for the RF trained on centrals and satellites together. We performed a $\chi^2$ test on these models and found our fiducial model captures clustering closer to TNG300 at the reliable scales (shaded region) with a score of 0.075, with the next best model, with a score of 0.105, being that of $q^2,\ \mathrm{vdisp},\ M_\mathrm{halo}$ where centrals and satellites are jointly trained. We emphasize, however, that these are noisy results and a larger volume which would result in more data points would be beneficial for accurate measurement. 

\section{Results from the emission line galaxy sample}
\label{appendix:ELGs}

In this appendix, we present the random forest (RF) results using the emission line galaxy sample (ELG-like) which are selected by color cuts and star-formation rates as described in \cite{HadTac20}. 

\begin{figure}
    \centering
    \includegraphics[width=0.9\columnwidth]{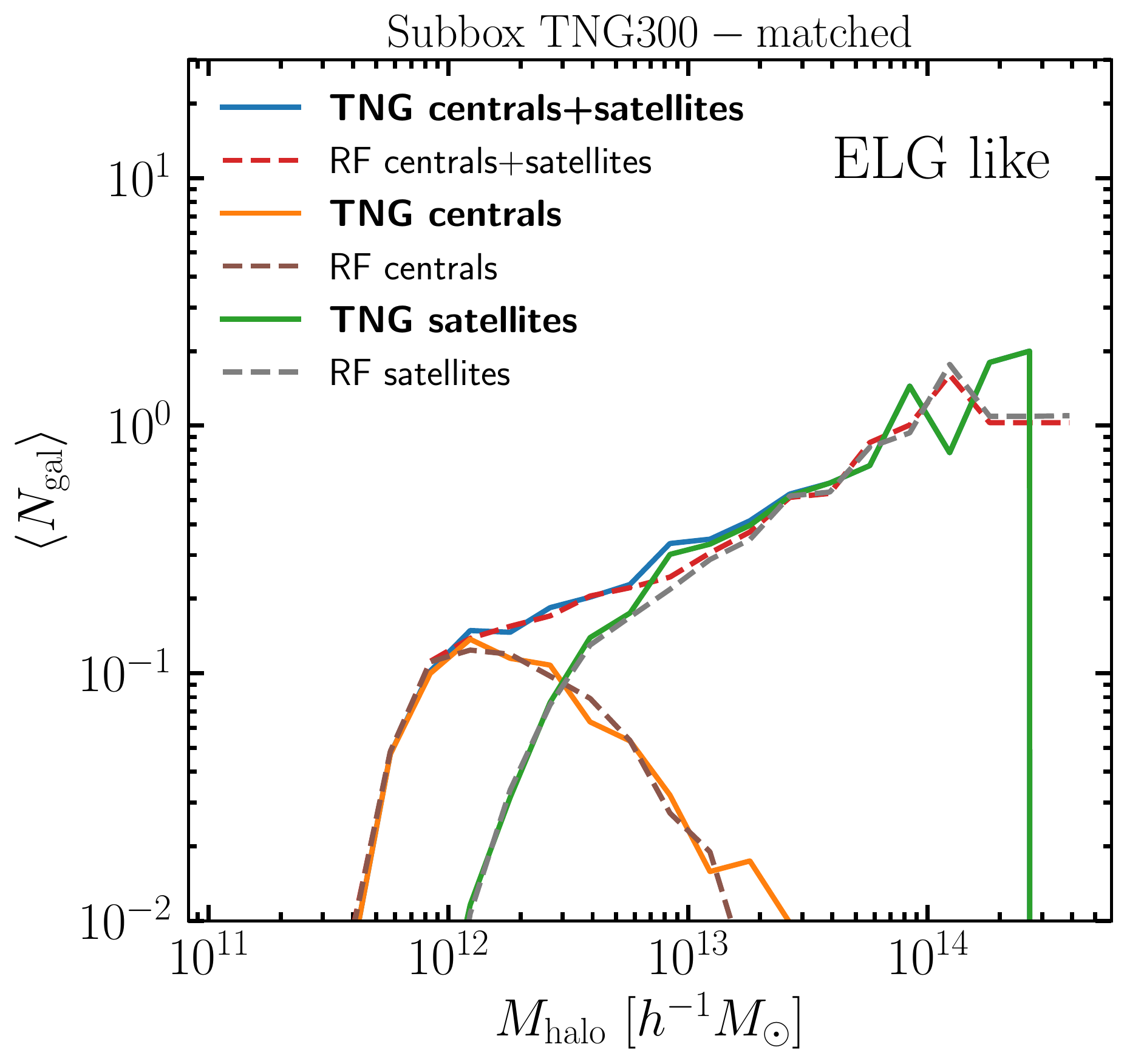}
    \caption{The mean emission line galaxy (ELG) occupation of TNG300 as a function of the halo mass. Similar to Fig. \ref{fig:hod_cent_sat}, we see that the random forest was also able to recover the mean occupation of ELG.}
    \label{fig:RF_HOD_ELG}
\end{figure}

\begin{figure}
    \centering
    \includegraphics[width=0.9\columnwidth]{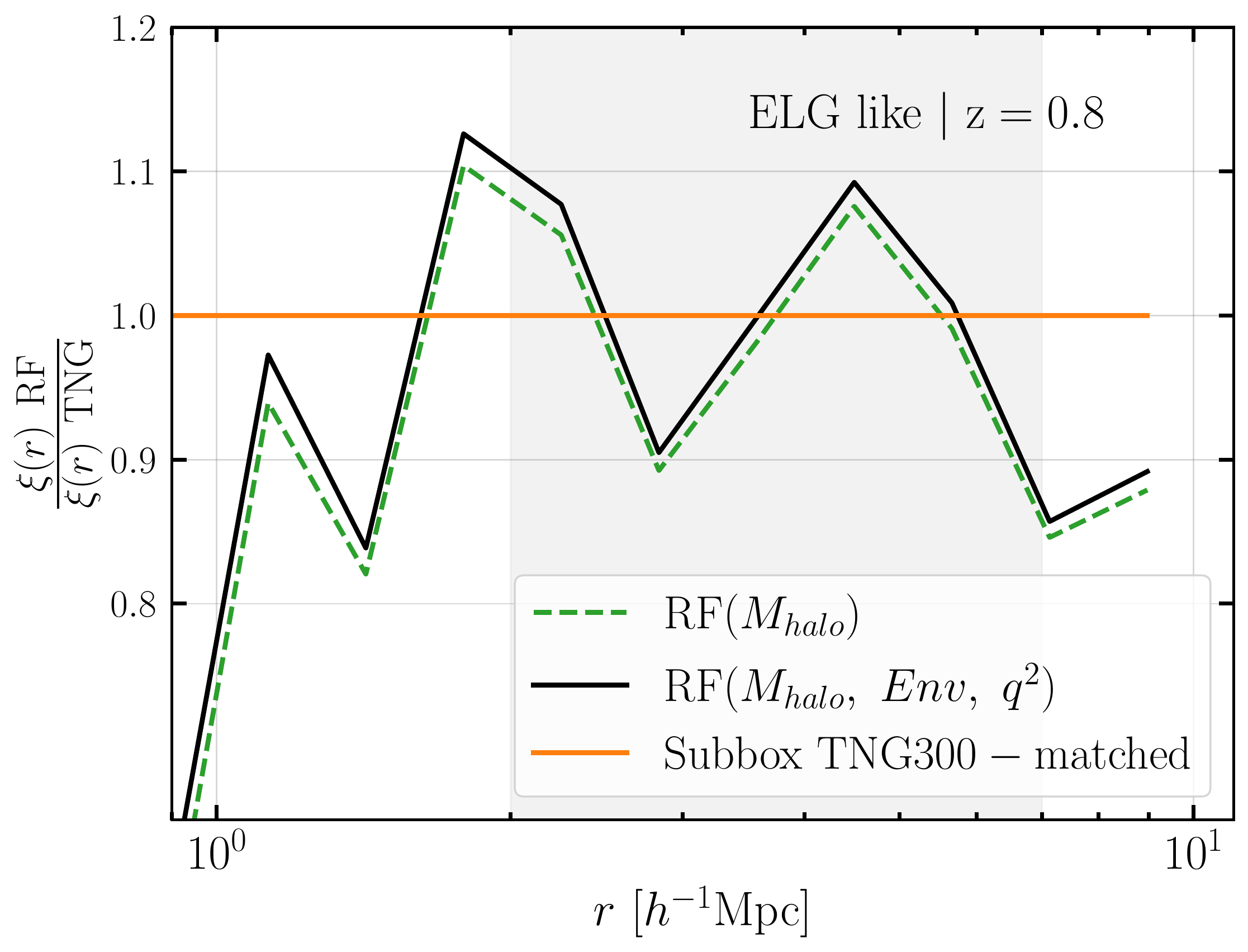}
    \includegraphics[width=0.9\columnwidth]{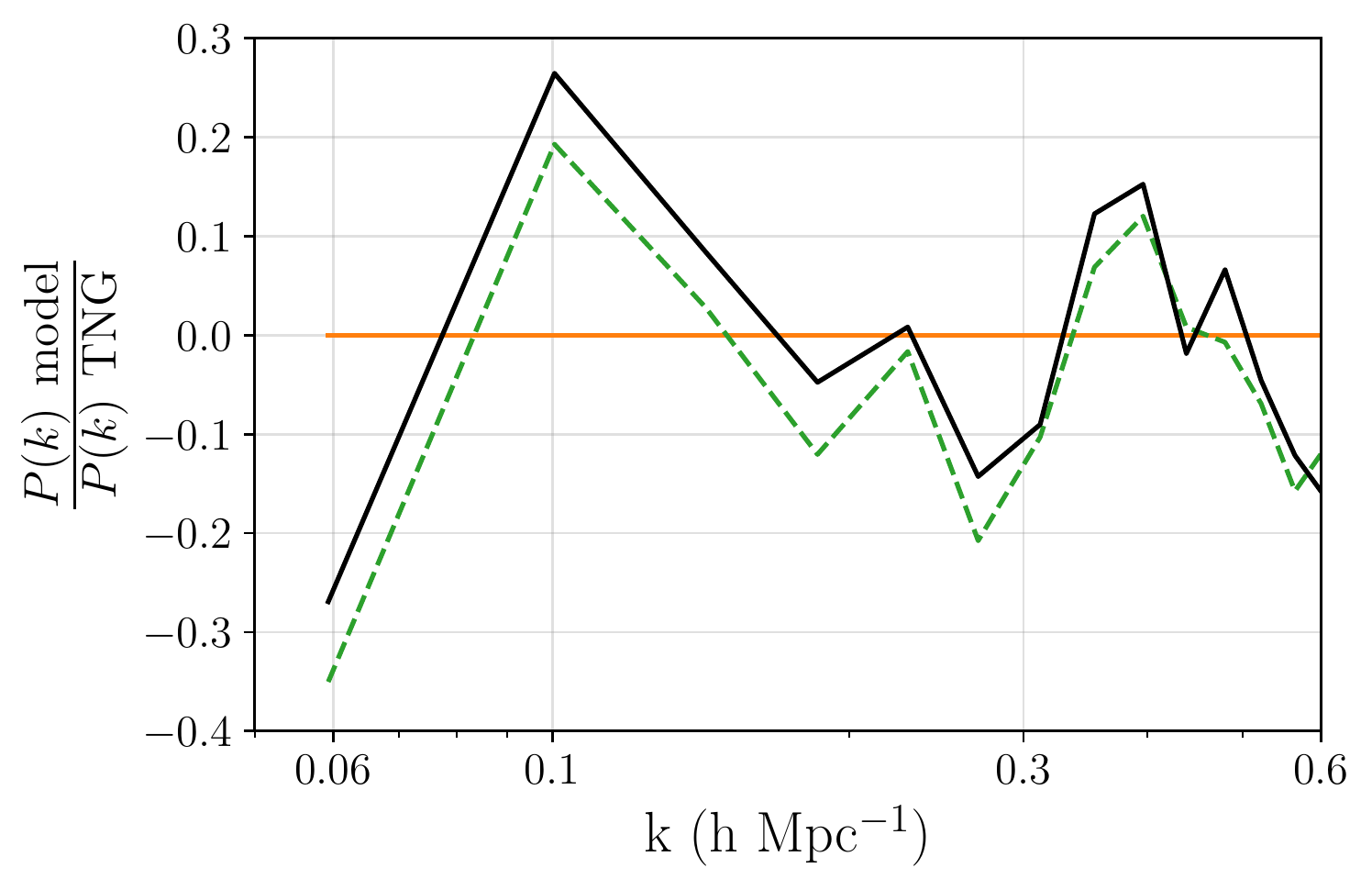}
    \caption{Same as Fig. \ref{fig:RF_results} but for emission line galaxies (ELG). We see that there is little difference between a mass only model and one incorporating secondary halo parameters.}
    \label{fig:RF_ELG}
\end{figure}

 It is important to note that the HOD looks different for ELGs,  shown in Fig. \ref{fig:RF_HOD_ELG}, compared to LRGs, shown in Fig. \ref{fig:hod_cent_sat}. We also see in Fig. \ref{fig:RF_HOD_ELG} that the RF was able to recover the galaxy occupations for the ELG-like sample. However, Fig. \ref{fig:RF_ELG} reveals that incorporating secondary halo parameters has little effect on the clustering for ELGs, consistent with previous studies \citep{HadTac20}; there is a significantly smaller discrepancy ($\sim$~3\%) from the mass-only model from ELGs as extracted from TNG300. The measurements of the correlation function are also quite noisy because of the small number of available ELGs. We again state the importance of a more comprehensive study of ELGs in future works with better statistics.

\section{Supplementary occupation statistics}

\subsection{Additional dependence of $N_{\mathrm{sat}}$ on secondary parameter}

\begin{figure}
\includegraphics[width=0.9\columnwidth]{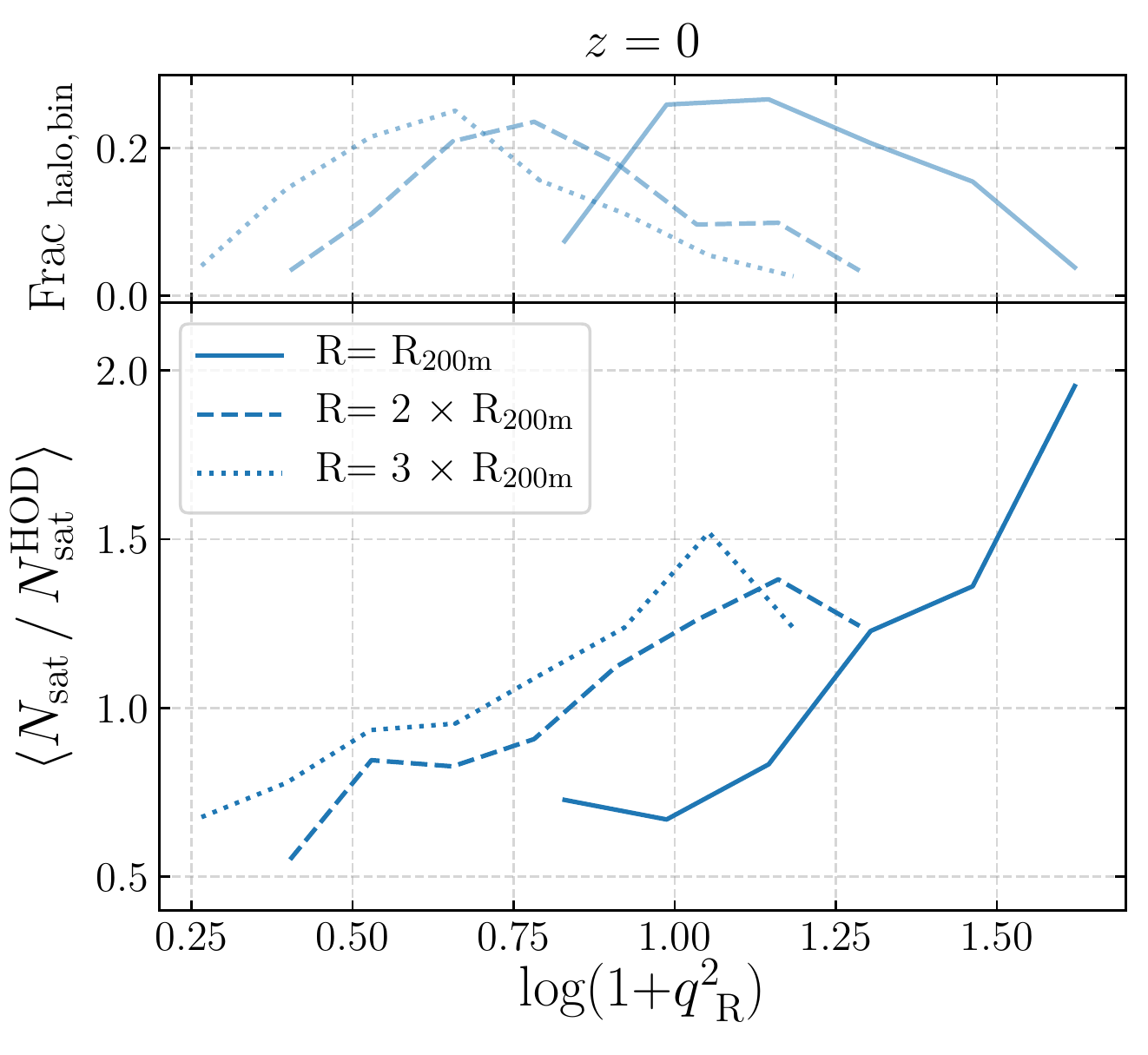}
\caption{Same as the left panel of Fig.~\ref{fig:Nsat_secondary} but showing the dependence on the radius at which the shear is calculated.}

\label{fig:shear_radius}
\end{figure}

Fig. \ref{fig:shear_radius} compliments Fig. \ref{fig:Nsat_secondary} by showing the dependence of the number of satellites on the radius at which the shear is calculated. 

\subsection{Modeling the HOD scatter}

We take advantage of the RF results to try and model the HOD scatter. We estimate the scatter as the standard deviation of the mean number of galaxies per mass bin, and compare $\sigma\langle N_{\mathrm{RF}} \rangle/\sigma\langle  N_{\mathrm{TNG}} \rangle$, where $N_{\mathrm{RF}}$ are the number of galaxies per mass bin predicted by the random forest, $N_{\mathrm{TNG}}$ are the number of galaxies per mass bin according to TNG300 and $\sigma$ is the standard deviation defined as

\beq
\sigma_{\rm Ngal}\equiv \sqrt{\frac{1}{N_{\rm halos}} \sum_i^{N_{\rm halos}}\Big(N^{i-\textup{true}}_\textup{gal}- N^{i-\textup{predicted}}_\textup{gal}\Big)^2}
\label{eq:STD}
\eeq

\begin{figure}
    \centering
    \includegraphics[width=0.9\columnwidth]{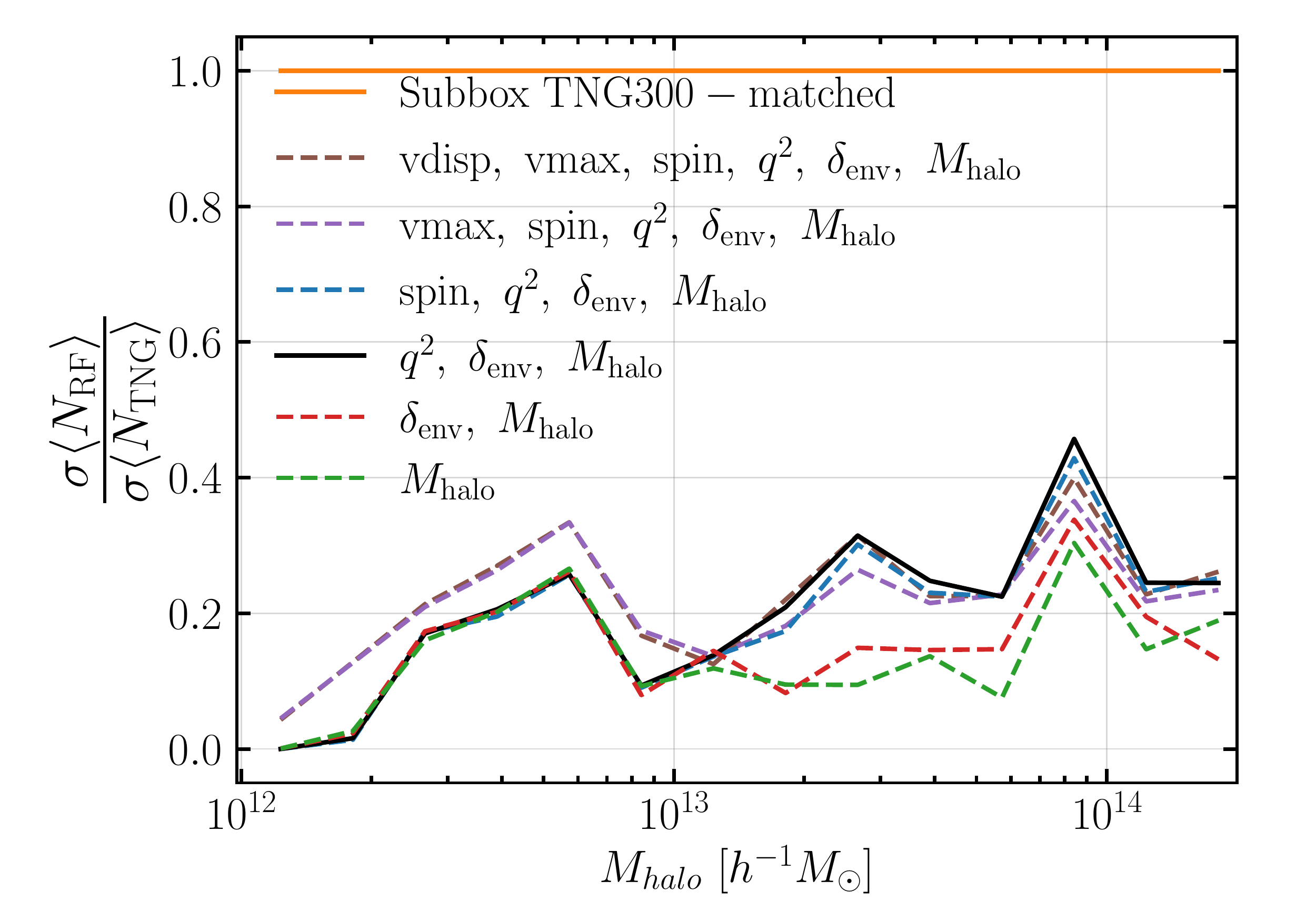}
    \caption{The HOD scatter predicted by the RF as compared to TNG300. The mass only model (green dashed line) is only able to capture $\sim$15\% of the scatter. By incorporating secondary halo parameters, environmental overdensity and shear (solid black line), the RF was able to capture $\sim$30\% of the scatter for halos with mass $>10^{13}h^{-1}M_{\odot}$.}
    \label{fig:RF_scatter}
\end{figure}
Fig. \ref{fig:RF_scatter} conveys that mass (green dashed line) is a poor predictor of the HOD scatter as the RF was only able to capture $\sim$15\% of the scatter using mass alone as a feature. While we did not obtain useful results by incorporating secondary halo properties, there were some minor improvements. There seems to be an interesting transition at $\sim10^{13}h^{-1}M_{\odot}$. We see that at lower masses, models that include Vmax (described in section \ref{Method:halo_properties}) captured about 5\% more of the scatter than mass alone, while all other models only perform as well as the mass only model. However at masses $>10^{13}h^{-1}M_{\odot}$ models that include shear performed better, with our preferred model, including environmental overdensity and shear (black solid line), capturing $\sim$30\% of the scatter. 


\bsp	
\label{lastpage}
\end{document}